\documentclass[draft]{agujournal2019}
\usepackage{url} 
\usepackage{lineno}

\usepackage{bbold}
\usepackage{amsmath}
\usepackage{amssymb}
\usepackage{comment}
\newcommand{\norm}[1]{\left \lVert #1 \right \rVert^2}
\newcommand{\normsqrt}[1]{\left \lVert #1 \right \rVert_2}
\usepackage{multirow} 
\usepackage{pifont}
\usepackage{verbatim}
\usepackage{booktabs}

\usepackage[finalnew]{trackchanges} 
\usepackage{soul}
\usepackage[T1]{fontenc}

\soulregister{\ref}{7}
\soulregister{\cite}{7}
\soulregister{\citeA}{7}
\soulregister{\eqref}{7}

\definecolor{royalpurple}{RGB}{188,66,245}

\newcommand{\cmark}{\ding{51}}  
\newcommand{\xmark}{\ding{55}}  
\newcommand{\diffmodel}[1]{\mathcal{M}_{#1}}
\newcommand{\obsmodel}{\mathcal{H}}
\newcommand{\lrdim}{d}
\newcommand{\hrdim}{n}
\newcommand{\sdedittime}{t_i}
\newcommand{\sdeditic}{x_i}

\definecolor{pptgreen}{RGB}{0,176,80}

\draftfalse

\journalname{Journal of Advances in Modeling Earth Systems (JAMES)}

\begin{document}

%
%


\title{Guided Unconditional and Conditional \\ Generative Models for Super-Resolution and Inference\\ of Quasi-Geostrophic Turbulence}

%
%




\authors{Anantha Narayanan Suresh Babu\affil{1,2}, Akhil Sadam\affil{2}, Pierre F.J. Lermusiaux\affil{1,2}}


\affiliation{1}{Department of Mechanical Engineering, Massachusetts Institute of Technology, Cambridge, USA}
\affiliation{2}{\mbox{Center for Computational Science and Engineering, Massachusetts Institute of Technology, Cambridge, USA}}




\correspondingauthor{Pierre F.J. Lermusiaux}{pierrel@mit.edu}



\begin{keypoints}
\item We benchmark four diffusion models for sparse super-resolution and inference of forced 2D quasi-geostrophic turbulence in the $\beta$-plane
\item Conditional diffusion models remain stable, predict correct statistics, and generalize even with sparse and gappy coarse observations
\item Guided unconditional models may ineffectively propagate information to unobserved regions when observation operators are sparse and local
\end{keypoints}

%
%

%
%


\begin{abstract}
Typically, numerical simulations of Earth systems are coarse, and Earth observations are sparse and gappy. We apply four generative diffusion modeling approaches 
to super-resolution and inference of forced two-dimensional quasi-geostrophic turbulence on the $\beta$-plane from coarse, sparse, and gappy observations. Two guided approaches minimally adapt a pre-trained unconditional model: SDEdit modifies the initial condition, and Diffusion Posterior Sampling (DPS) modifies the reverse diffusion process's score. Two conditional approaches, a vanilla variant and classifier-free guidance, require training with paired high-resolution and observation data. 
We consider multiple test cases spanning: two regimes, eddy and anisotropic-jet turbulence; two Reynolds numbers, $10^3$ and $10^4$; and two observation types, $4 \times$ coarse-resolution fields and coarse, sparse and gappy observations. Our comprehensive skill metrics include norms of the reconstructed vorticity fields, turbulence statistical quantities, and 
quantifications of the super-resolved probabilistic ensembles and their errors. We also study the sensitivity to tuning parameters such as guidance strength. 
Results show that the generated super-resolution fields of
SDEdit are unphysical, 
while those of DPS are reasonable but with smoothed fine-scale features; however, neither of these lower-cost models
propagates observational information effectively to unobserved regions.
The two conditional models require re-training, but 
reconstruct missing fine-scale features, are cycle-consistent with observations, and predict correct turbulence statistics, including the tails. Further, their mean errors are highly correlated with and predictable from their ensemble standard deviations. Results highlight the trade-offs 
between ease of implementation, fidelity (sharpness), and cycle-consistency of the diffusion models, and offer practical guidance for deployment in geophysical inverse problems.
\end{abstract}

\section*{Plain Language Summary}

Simulating the ocean or the atmosphere at high resolution is expensive because computers solve physics equations on many tiny grid cells. As an approximation, simulations are run with a small number of large grid cells, missing important small-scale physics such as swirling eddies. Even geophysical observations with satellites, balloons, buoys, or robots can be incomplete with large gaps. We explore whether generative diffusion models, a new machine learning technique, can complete the missing details in ways that respect the laws of physics. Using turbulence simulations, we compare four diffusion models. Two \emph{conditional} models require training from scratch to directly learn missing details, demanding more data and computation, while two \emph{guided unconditional} models modify existing diffusion models to complete missing details. We find that conditional models create realistic details that respect turbulence physics. In challenging cases, one of the unconditional models fails, while the other can provide low-cost, reasonable guesses. We show tradeoffs between resource use and realism that users can tune to their needs. Some generative diffusion models can lead to sharper maps of ocean currents, weather, and climate, help merge real observations with simulations, and possibly learn physics from even incomplete, indirect data.

\section{Introduction}

Higher-resolution oceanic and atmospheric modeling for marine, weather, and climate 
predictions \cite{kalnay2003atmospheric,
harris2014global,
chassignet2021importance,
brotzge2023challenges,
lai2024machine}
can provide greater accuracy but are computationally expensive. 
Numerical models are limited not only by their spatio-temporal resolution but also by the processes they represent and the number of variables they simulate \cite{gupta_lermusiaux_PRSA2021,jacobs2023adapting}. For example, simulating the full range of spatio-temporal scales in turbulent flows is computationally intractable; 
approximations and parameterizations, such as in Large-Eddy Simulations (LES), are thus employed to represent the unresolved scales, which can also limit utility \cite{leonard1975energy, pope2001turbulent}. Recent machine-learning-based surrogate models for the ocean and atmosphere \cite{pathak2022fourcastnet,rajagopal_et_al_Oceans2023} also contain unresolved scales due to spectral bias, which inhibits accurate prediction of fine-scale features \cite{lai2024machine}. This can be amplified by dynamics such as quasi-geostrophic turbulence, leading to unstable or nonphysical long-term predictions \cite{chattopadhyay2023long}. The dual-cascade of forced geophysical turbulence \cite{kraichnan1971inertial}, i.e., the inverse cascade of energy from forcing scales to large scales and the forward cascade of enstrophy to small scales, is a challenge. 
In the ocean, resolving submesoscale processes with scales of the order of $100$\;m to $10$\;km \cite{mcwilliams2016submesoscale, taylor2023submesoscale} can contribute to energy, mass, and tracer transports \cite{thomas2008submesoscale}. In climate, simulations with scales of $10$\;km can improve predictions of extreme weather events and enhance risk assessment \cite{diffenbaugh2005fine, lopez2025dynamical}. 
As numerical models, in situ observations from weather stations, balloons, buoys, drifters, gliders,  and ships, as well as remote observations from satellites and radars, are also spatio-temporally sparse and gappy \cite{lin2020ocean,bluestein2022atmospheric}.
The integration of modeling and observing systems for intelligent sampling and data assimilation can be most useful to improve estimates \cite{bennett2005inverse,lermusiaux_et_al_O2006a,
kalnay2003atmospheric,law2015data,lermusiaux_et_al_TheSea2017}.
In general, techniques that reconstruct fine-scale features from coarse or sparse simulated or observed data are critical to improve their utility \cite{buzzicotti2023data}. One such approach to reconstruct fine-scale features is through super-resolution \cite{sofos2025review}, which is the focus of our work. 

Super-resolution can be interpreted as a framework for a broad class of inverse problems \cite{fukami2023super}, with applications to several Earth system modeling challenges:
\\
\noindent
\emph{1.\ Subgrid-scale (SGS) modeling}. 
    A goal of model closure is to parametrize unresolved subgrid-scale processes. Traditional approaches include functional models, often based on eddy viscosity hypotheses \cite{smagorinsky1963general, leith1968diffusion} and structural models, which directly approximate the subgrid-scale stress terms through scale-similarity or formal series expansions \cite{clark1979evaluation,bardina1980improved}. Purely diffusive functional approaches do not account for backscatter, the transfer of energy from subgrid scales to resolved scales 
    \change{ \cite{grooms2023backscatter}}{\cite{frederiksen1997eddy, jansen2014parameterizing}}, and are not well correlated with the true SGS term, while structural models often suffer from numerical instabilities \cite{vreman1996large}. 
    To overcome these issues, recent data-driven closures have been proposed \cite{maulik2019subgrid, zanna2020data, gupta_lermusiaux_SR2023, ross2023benchmarking, jakhar2024learning}. 
    Super-resolution provides an alternate framework to develop 
    optimal 
    closures by reconstructing subgrid-scale features \cite{langford1999optimal,bode2021using,maejima2025unsupervised}. 
\\
\noindent    
    \emph{2.\ Downscaling}.
    Downscaling aims to improve the fidelity of regional and shorter-term forecasts by increasing the spatio-temporal resolution of coarser global Earth system models. Dynamical downscaling and nesting \cite{von2000spectral,denis2002downscaling,haley_lermusiaux_OD2010,kulkarni_et_al_Oceans2018,johnston_et_al_Oceanog2019b,simmons2019dynamical} use global model outputs to provide large-scale boundary conditions, initial conditions, and forcing to regional model simulations. These schemes have varied accuracies and computational costs, e.g., \cite{xu2019dynamical}.
    With statistical downscaling, statistical relationships are learned between historical regional data and model outputs through regression or stochastic generation \cite{wilby1998statistical, khan2006uncertainty}. This statistical approach can downscale larger ensembles due to their efficiency and reduce biases inherent to global models. 
    Some applications can also favor specific downscaling, e.g., \cite{keller2022downscaling}.
    Recently, machine learning has emerged as a promising tool for statistical downscaling \cite{bano2020configuration, rampal2024enhancing}, with ensemble super-resolution offering an advanced nonlinear approach. These super-resolved fields still require bias correction to reduce systematic errors \cite{wan2023debias}.  
\\
\noindent
    \emph{3.\ State estimation and mapping}. 
    State estimation is the process of determining the spatial and temporal distribution or evolution of states (dynamical fields).
    Optimal interpolation (or objective mapping), estimates states from observations through statistical interpolation \cite{bretherton1976technique, menemenlis1997adaptation}. 
    These estimates, however, may not obey the governing dynamics. 
    Data assimilation instead melds observations with the underlying dynamics from numerical models
    \cite{robinson_et_al_Sea1998,
    robinson_lermusiaux_EOS2001,law2015data,carrassi2018data,hoteit2018data}. In oceanic and atmospheric modeling, popular data assimilation methods are based on Kalman filtering and its ensemble and non-Gaussian extensions \cite{evensen2003ensemble, sondergaard_lermusiaux_MWR2013_part1}, particle filters \cite{van2019particle}, and variational methods such as 3D and 4D-VAR \cite{bannister2017review}. Recently, machine learning models have been developed for data assimilation \cite{bocquet2023surrogate, hodyss2026using}, where data-driven surrogates replace either the dynamical model \cite{adrian2025data} or the entire end-to-end data assimilation \cite{manshausen2024generative, martin2025generative}. 
    Super-resolution can also be used to reconstruct missing features in physical space, with additions to preserve dynamics \cite{barthelemy2022super}. Unlike subgrid scale modeling or downscaling, where missing features typically correspond to high wavenumbers in Fourier space, here reconstructed features also correspond to low wavenumbers \cite{buzzicotti2023data}.
\\
\noindent
    \emph{4.\ Scale-enhanced initialization and simulation spin-up}. 
    In geophysical modeling, interpolation across grids 
    is utilized to initialize and spin-up regional models from global analyses.
    For example, to initialize coastal ocean forecasts, global fields can be adjusted to high-resolution bathymetry and multiscale dynamics using data corrections and
    least-square inversions for multi-process adjustments \cite{haley_et_al_OM2015,lermusiaux_et_al_Oceans2024}.
    High-resolution large-ensemble forecasts can be initialized by generating multi-region, downscaled, three-dimensional field perturbations based on data and dynamics \cite{lermusiaux_et_al_QJRMS2000,lermusiaux_JAOT2002,lermusiaux_et_al_BBN_Oceans2020,haley_et_al_Oceans2023}.
    As in dynamical downscaling, larger-scale information can also be re-injected into the regional model to provide a warm-start \cite{short2022reducing}.
    As all of these scale-enhanced initializations take computational time,
    fast super-resolution would be most useful to generate multiscale-consistent ensemble initializations or enable direct high-resolution spin-up through super-resolved warm starts.
    
Given the wide application range, super-resolution of turbulent flows is a growing area of research in machine learning. Various data-driven approaches and deep neural architectures have been explored, including sparse representations \cite{callaham2019robust}, convolutional neural networks (CNNs) with hybrid downsampled skip-connection/multi-scale networks (DSC/MS) \cite{fukami2019super, fukami2024single}, spatio-temporal CNNs \cite{fukami2021machine, liu2020deep}, physics-informed neural networks (PINNs) \cite{gao2021super}, graph neural networks (GNNs) \cite{barwey2025mesh}, vision transformers \cite{xu2023super},
and fully differentiable solvers that train with a dynamics-based loss \cite{page2025super}. While these methods can generate plausible high-resolution fields, they provide only a single best estimate without uncertainty quantification and often struggle to preserve physical invariants like energy spectra.
Since nonlinear and high-dimensional inverse problems suffer from the \emph{curse of dimensionality} and are inherently ill-posed \cite{fernandez2020curse}, there is a growing need for probabilistic methods that generate samples with uncertainty quantification \cite{lermusiaux_et_al_O2006b}. 

Probabilistic generative models provide a natural framework to meet these requirements. Popular generative models include variational autoencoders (VAEs), generative adversarial networks (GANs), normalizing flows (NFs), and diffusion models (DMs). VAEs 
map fields to samples from a low-dimensional latent prior, typically Gaussian in distribution \cite{kingma2019introduction}. 
Although vanilla VAEs generate blurry samples \cite{bredell2023explicitly},
they have been used for super-resolution through modal decomposition \cite{eivazi2022towards}.
In contrast, GANs generate sharp samples by matching distributions \cite{gui2021review}, showing effectiveness for some inverse problems \cite{patel2022solution, guemes2022super}. 
However, GAN training is known to be unstable and susceptible to mode collapse, as it requires the simultaneous training of two competing neural networks: a generator network that generates plausible samples to fool a discriminator network \cite{wiatrak2019stabilizing, kossale2022mode}. Vanilla GANs are potentially inaccurate, so \citeA{kim2021unsupervised} utilize cycle-consistent-GANs \cite{zhu2017unpaired} for super-resolution of Direct Numerical Simulation fields from LES channel flows.
Finally, normalizing flows (NFs) are stable to train and produce sharp images; they learn an \emph{invertible} map to a tractable distribution \cite{rezende2015variational}. These models excel at density estimation, but were historically limited in architecture by the invertability constraint \cite{zhai2024normalizing}. Recent extensions include continuous time flows through flow matching and stochastic interpolants \cite{lipman2022flow, albergo2022building}, with applications to super-resolution of Kolmogorov flows \cite{chen2024probabilistic}.

Diffusion models have emerged as state-of-the-art generative tools \cite{song2020score}. 
Diffusion models are more stable to train than GANs and have been shown to generate fields of equivalent or higher quality than GANs, VAEs, or NFs \cite{dhariwal2021diffusion}. These models first apply a corruption process that destroys structure in the fields from a given distribution, and then train a model that iteratively reverses this corruption process to generate fields that follow the required distribution. The reverse process could be thought of as iterative spectral regression, where the large-scale features are predicted first, with small-scale features added iteratively \cite{wang2023diffusion}. Diffusion models are flexible and can be either conditional \cite{saharia2022palette}, where the model is directly trained to learn the relationship between the input and the target, or guided unconditional \cite{daras2024survey}, where a pre-trained model is steered towards the target using a separate guidance process. 

Due to these properties, generative diffusion models have been applied to inverse problems in computer vision and geophysical fluid dynamics \cite{milanfar2025denoising}. 
For super-resolution, most approaches rely on conditional diffusion models \cite{sardar2024spectrally, oommen2025integrating} though guided unconditional models also exist \cite{fan2025neural} with applications to idealized Kolmogorov flows. 
For downscaling of atmospheric flows, some approaches have conditioned diffusion models on neural surrogates \cite{sundar2024taudiff, mardani2025residual} or directly on coarse-resolution model output \cite{watt2024generative, han2025diffusion}. \citeA{chakraborty2025multimodal} guided an unconditional model for generative atmospheric flow downscaling.
For state estimation and inference, guided unconditional models have been employed for generative data assimilation of sparse observations in idealized turbulent flows \cite{rozet2023score} and weather fields \cite{manshausen2024generative}. Unconditional \cite{martin2025generative} and conditional diffusion models \cite{li2023multi, asefi2025generative, souza2025surface} have been used to infer surface and subsurface ocean and idealized vorticity fields from coarse, sparse, and gappy observations. To reduce training costs, training-free diffusion approaches based on ensemble filtering have also been proposed \cite{yin2024scalable, bao2025nonlinear}.

Despite these advances, no study has focused on both super-resolution and estimation or inference of geophysical turbulent fields from coarse full field and coarse, sparse and gappy data. 
Moreover, existing works focus on a single method in isolation without direct comparison across methods, with several open questions, including: While guided unconditional models are effective in computer vision, can they achieve comparable skill to conditional models in geophysical fluid super-resolution problems? When do the assumptions made in guided unconditional models break down in turbulent flow settings? Can conditional models remain stable \cite{baldassari2023conditional} and generalize effectively when observations are sparse and gappy? How does observational information propagate in diffusion models? How sensitive are outputs of diffusion models to the guidance strength (or other tuning parameters)?

To address these challenges, (1) We apply four diffusion modeling (2 guided unconditional and 2 conditional) approaches to super-resolution and inference of quasi-geostrophic turbulence \emph{under the $\beta$-plane approximation} for different Reynolds numbers and flow regimes. (2) We investigate the effect of sparse and gappy coarse observations. (3) We benchmark the four score-based diffusion models using a comprehensive suite of skill metrics, including norms of the generated vorticity fields and turbulence statistics. (4) We quantify and evaluate the uncertainty of our generative models
using the generated ensemble of super-resolved fields
and their errors. (5) We analyze the propagation of sparse and gappy observational information by the generative diffusion processes. (6) We study the sensitivity of the four diffusion models to their guidance strength. Our results highlight the trade-offs between ease of implementation, fidelity (sharpness), and cycle-consistency when deploying diffusion models for super-resolution and inference applications in geophysical fluid dynamics \cite{bennett1992inverse,wunsch2006discrete}.

The paper is organized as follows: Sect.\;(\ref{sec: PS}) introduces the formal problem statement for super-resolution and state estimation or inference. Details of the various guided unconditional and conditional score-based diffusion modeling methods utilized are in Sect.\;(\ref{sec: diffusion}). Sect.\;(\ref{sec: Methods}) describes the quasi-geostrophic turbulence simulations used in our applications, along with details of the dynamical regimes, numerical schemes, datasets, test cases, and skill metrics. Applications of the four diffusion model methods to super-resolution and inference of quasi-geostrophic turbulence with coarse and with sparse, gappy observations are provided in Sect.\;(\ref{sec: results}), followed by a discussion of the comparative analyses and conclusion (Sect.\;\ref{sec: conclusions}).

\section{Problem Statement} \label{sec: PS}

We consider two related inverse problems: 1) super-resolution \cite{moser2023hitchhiker}
where missing short-wavelength, subgrid-scale features are reconstructed, and 2) state estimation or inference, where high-resolution fields are estimated from low-resolution, sparse, and gappy data, with reconstructed features including long wavelengths. We formulate both problems within a common framework as an inversion of an observation process.

Specifically, we define a measurement model that maps a high-fidelity field $x_0 \in \mathbb{R}^{\hrdim}$ to a low-fidelity field $y \in \mathbb{R}^{\lrdim}$ (the low-resolution data), where $\lrdim \leq \hrdim$, through a nonlinear observation operator $\obsmodel: \mathbb{R}^{\hrdim} \to \mathbb{R}^{\lrdim}$, where $z\sim \mathcal{N}(0,\mathbb{I}_{\lrdim})$ and $\sigma_y$ models the noise in the observation process.
\begin{equation}\label{eq:prob_state}
y = \obsmodel(x_0) +\sigma_y z
\end{equation}
In general, the observation operator is not restricted to just the physical space, but could also act in the Fourier domain (wavenumber space). Observations can also be partial, with only a subset of the domain being observed. We are interested in the inverse of \eqref{eq:prob_state}, or a model~$\diffmodel{}$ that, given noisy observation samples $y$, can generate plausible reconstructed high-fidelity samples $\hat{x}_0$ from the conditional distribution $p(x_0 | y)$. Because this inverse mapping is not one-to-one, it is ill-posed, and perfect recovery is impossible for many applications in geophysical modeling \cite{bennett1992inverse, blau2018perception}. Therefore, we seek to obtain generative models $\diffmodel{}$ that perform such inversions in a probabilistic sense, and sample from the posterior distribution rather than provide only point estimates such as the conditional expectation $\mathbb{E}[x_0|y]$. We propose using generative diffusion models due to their ability to effectively sample an ensemble of plausible fields that follow a required distribution \cite{dhariwal2021diffusion}.

This problem formulation is related to data assimilation \add{\cite{law2015data, evensen2022data}.}\change{, in particular to}{ For example, early approaches such as optimal interpolation (OI) \cite{gandin1963objective} or} 3D-VAR \cite{courtier1998ecmwf}\remove{, or equivalently to, since it} 
\add{also}
perform\remove{s} inversions one snapshot at a time\remove{.} \add{and optimize for a single analysis, i.e., the minimum of the error variance or of a cost function, respectively \cite{lorenc1986analysis,xie2002impact}. Under assumptions of Gaussianity, OI and 3D-VAR correspond to the maximum a posteriori (MAP) estimate of the posterior, but neither scheme directly provides dynamic uncertainty quantification \cite{kalnay2003atmospheric}. Ensemble-based methods utilize an ensemble of numerical model forecasts to evolve and predict uncertainties, and  include ensemble Kalman filters, error subspace schemes, sequential Monte Carlo schemes, and particle filters \cite{evensen2003ensemble,van2019particle}. For nonlinear data assimilation, several of these schemes utilize the forecast non-Gaussian distributions and semiparametric model fits such as Gaussian Mixture Models \cite{sondergaard_lermusiaux_MWR2013_part1,lolla_lermusiaux_MWR2017_partII,
carrassi_etal_DA_chaatic_2022,evensen2022data}.
However, the accuracy of ensemble-based methods is limited by finite subspace and sampling sizes, often constrained by computational cost, which can lead to sampling errors or degeneracy in high-dimensional settings \cite{ehrendorfer2007review, bengtsson2008curse}. Initializing ensembles with realistic uncertainty structures also requires data and modeling \cite{lermusiaux_et_al_QJRMS2000,lermusiaux_JAOT2002,
zupanski2006initiation}. In contrast, trained diffusion-based models can generate larger observation-consistent ensembles at much lower computational cost \cite{manshausen2024generative}. Rather than relying on numerical forecast ensembles, 
a generative diffusion model could provide a learned prior with a 
typically non-Gaussian distribution of plausible high-resolution fields $p(x_0)$. Of course, training such models could be costly, requiring significant data and long numerical 
simulations. In addition, retraining and targeted denoisers may be needed at assimilation times for conditioning and optimal dynamic prior distributions.
Several of these diffusion-based assimilation limitations are discussed by \citeA{hodyss2026using}.
In this work, we evaluate and compare the accuracy, capabilities, and costs of four generative diffusion models for super-resolution and inference of turbulent fields in the sense of \eqref{eq:prob_state}.} 

\remove{Two key distinctions arise.
First, in 3D-VAR, the prior (or background state) is commonly supplied by a numerical model, whereas here it is provided by a trained
diffusion model that has learned a generative prior \cite{manshausen2024generative}, i.e., the distribution of all possible high-resolution fields $p(x_0)$, which is in general not Gaussian. Second, while classic 3D-VAR optimizes for a single deterministic analysis \cite{xie2002impact}, i.e., the minimum of a cost function, which corresponds to the maximum a posteriori (MAP) estimate under assumptions of Gaussianity, our diffusion-based approach can produce observation-consistent ensembles.}

In summary, our goal is to solve the inverse problem of reconstructing high-resolution vorticity fields in quasi-geostrophic turbulence under the $\beta$-plane approximation (Sect.\;\ref{sec: Methods}) from coarse, sparse, and gappy observations using trained generative diffusion models that sample from the posterior distribution $p(x_0 | y)$. For our focus on super-resolution, $\obsmodel$ corresponds to filtering and down-sampling operations. For super-resolution with inference, this is combined with sparse and gappy observing systems. 
The specific forms of $\obsmodel$ are detailed in Sect.\,(\ref{sec: filtering}).

\section{Score-Based Diffusion Models} \label{sec: diffusion}

We now describe the score-based generative diffusion models and their use for super-resolution and inference. Diffusion models aim to generate samples from a complex distribution $p_0$ (e.g., the distribution of all possible fields that can be simulated using the Quasi-Geostrophic equations) using a more tractable distribution $p_T$ (e.g., a multivariate Gaussian distribution). 
This generative process occurs over a pseudo-time or diffusion time $t$ within the interval $[0,T]$ where $T$ is the maximum diffusion time. This pseudo-time is the artificial time variable of the generative sampling process and is not a physical time. 
The use of diffusion models typically involves a forward process, which consists of gradually \emph{noising} or corrupting samples from $p_0$ to resemble samples of $p_T$ over pseudo-time, and a reverse process, which consists of solving a deterministic or stochastic differential equation (SDE) \cite{song2020score} or a discrete Markov-chain process \cite{sun2022score}, starting from the tractable distribution $p_T$ to generate samples from $p_0$. This generation process can be conditioned on some constraints, and the sampling can be very efficient on modern computers \cite{sohl2015deep}. 
Following \citeA{song2020score}, the forward process for a diffusion model can be represented by an Îto differential equation
\begin{equation} \label{eq:ito_forward}
    dx_t = f_t(x_t)dt + g_t\,dW_t \qquad \forall \, t \in [0, T]
\end{equation}
with initial conditions $x_0 \sim p_0$, i.e., samples from the complex distribution $p_0$. In \eqref{eq:ito_forward}, $x_t \in \mathbb{R}^{\hrdim}$ is the field generated at intermediate pseudo-times $0\leq t \leq T$, $f$ the drift coefficient, $g$ the diffusion coefficient, and $W$ the standard Wiener process or Brownian motion. The subscript $t$ highlights the pseudo-time dependence. Typically, the forms of $f$ and $g$ are chosen a priori and hence known. This forward process has a corresponding stochastic reverse process that can be derived analytically utilizing results from stochastic calculus \cite{anderson1982reverse}, 
\begin{equation} \label{eq:ito_reverse}
    dx_t = \bigg[f_t(x_t) - g_t^2\nabla_{x_t}\bigg(\log p_t(x_t)\bigg)\bigg]\,dt + g_t\,dW_t
\end{equation}
where $\log p_t(x_t)$ is the log-probability distribution at pseudo-time $t$. Its gradient, $\nabla_{x_t} \log p_t(x_t)$, is referred to as the \emph{score}. It appears in the deterministic drift term of the reverse diffusion process and guides it, directing noisy or corrupted samples at pseudo-time $t$ towards regions of higher probability density of the distribution $p_0$ \cite{park2024random}. This score has no closed-form analytical expression and is typically estimated using a deep neural network or through training-free Monte Carlo approximations \cite{liu2024diffusion,bao2025nonlinear}. If instead we are interested in generating fields given certain constraints (e.g., high-resolution fields that correspond to a specific low-resolution field), the resulting reverse conditional diffusion process can be derived as
\begin{equation} \label{eq:ito_reverse_conditional}
    dx_t = \bigg[f_t(x_t) - g_t^2\nabla_{x_t}\bigg(\log p_t(x_t|y)\bigg)\bigg]\,dt + g_t\,dW_t
\end{equation}
where $y$ is the conditioning field and $\log p_t(x_t|y)$ is called the \emph{conditional score}.

In this work, we utilize the Variance-Preserving (VP-SDE) forward process \cite{song2020score}, where the marginal distribution at each step during the forward diffusion process \eqref{eq:ito_forward} is given by 
\begin{equation}\label{eq: forward process}
     p_t(x_t) = \mathcal{N}(x_t;\mu_t x_0,\sigma_t^2 \mathcal{I}_{\hrdim}),
      ~~~ 0 \leq \sigma_t \leq 1 \, , ~~~ \underset{t \to T}{\lim} \sigma_t \to 1
\end{equation}
where $x_0 \sim p_0$ denotes samples from the complex distribution $p_0$, the amplitude $\sigma_t$ defines the \emph{noise schedule}, and the normalization factor $\mu_t = \sqrt{1-\sigma_t^2}$ ensures the variance of noisy samples $x_t$ is preserved over pseudo-time $0\leq t \leq T$. 
This equation \eqref{eq: forward process} is the marginal of the analytical solution of \eqref{eq:ito_forward} under the VP-SDE assumption, and provides a computationally efficient method to generate noisy or corrupted samples at intermediate pseudo-times.
Using Tweedie's formula \cite{robbins1992empirical}, the score for the unconditional reverse process is then given by
\begin{equation} \label{eq: Tweedie unconditional}
\nabla_{x_t}\bigg(\log p_t(x_t)\bigg) = \frac{\mu_t \mathbb{E}[X_0 | X_t =x_t] - x_t}{\sigma_t^2}
\end{equation}
and the conditional reverse process is given by
\begin{equation} \label{eq: Tweedie conditional}
\nabla_{x_t}\bigg(\log p_t(x_t|y)\bigg) = \frac{\mu_t \mathbb{E}[X_0 | X_t =x_t, Y = y] - x_t}{\sigma_t^2}
\end{equation}
where $\mathbb{E}$ is the expectation operator, and   $\mathbb{E}[X_0 | X_t =x_t]$ and $\mathbb{E}[X_0 | X_t =x_t, Y = y]$ correspond to well-trained \emph{denoiser} neural networks, $h_{\theta}(.,t)$, in practice. 

For the unconditional reverse process, the \emph{denoiser} neural network $h_{\theta,\text{uncond.}}(x_t,t)$ with trainable parameters $\theta$ is trained to predict the noise added given a noisy field $x_t$ as the input. The neural network is trained to minimize the following objective:
\begin{equation} \label{eq: training_obj_uncond}
    \underset{\theta}{\text{min}}\, \underset{x_0,x_t}{\mathbb{E}} \norm{h_{\theta,\text{uncond.}}(x_t,t)-\frac{x_t - \mu_t x_0}{\sigma_t}}
\end{equation}
For the conditional models, the neural network  $h_{\theta,\text{cond.}}(x_t,y,t)$ also takes in the conditioning field ($y$, the observations) as an additional input, and hence the training objective is modified as
\begin{equation} \label{eq: training_obj_cond}
    \underset{\theta}{\text{min}}\, \underset{x_0,x_t,y}{\mathbb{E}} \norm{h_{\theta,\text{cond.}}(x_t,y,t)-\frac{x_t - \mu_t x_0}{\sigma_t}}
\end{equation}
The learned score, $s_{\theta}(.,t)$, for both conditional and unconditional models can be computed using their corresponding denoiser networks $h_{\theta}(.,t)$
\begin{equation} \label{eq: training_score}
    s_{\theta}(.,t) = -\frac{h_{\theta}(.,t)}{\sigma_t}
\end{equation}

Finally, for sampling, the reverse diffusion process for both unconditional \eqref{eq:ito_reverse} and conditional models \eqref{eq:ito_reverse_conditional} are discretized in time and integrated using an exponential integrator scheme \cite{zhang2022fast, song2020denoising} with Langevin Monte Carlo corrections \cite{song2020score, rozet2023score} using the  learned score \eqref{eq: training_score}. 

Next, we describe the four generative diffusion modeling approaches we employ and compare them inspired by the following questions: Is the score modified? Is prior knowledge of the down-sampling and filtering operations required? Is task-specific retraining required? Is the guidance 
controllable? Finally, we compare the number of function evaluations (neural network evaluations) per diffusion pseudo-timestep.

\subsection{Guided Unconditional Models} \label{sec: unconditional}

\begin{figure}
    \includegraphics[width=1\linewidth]{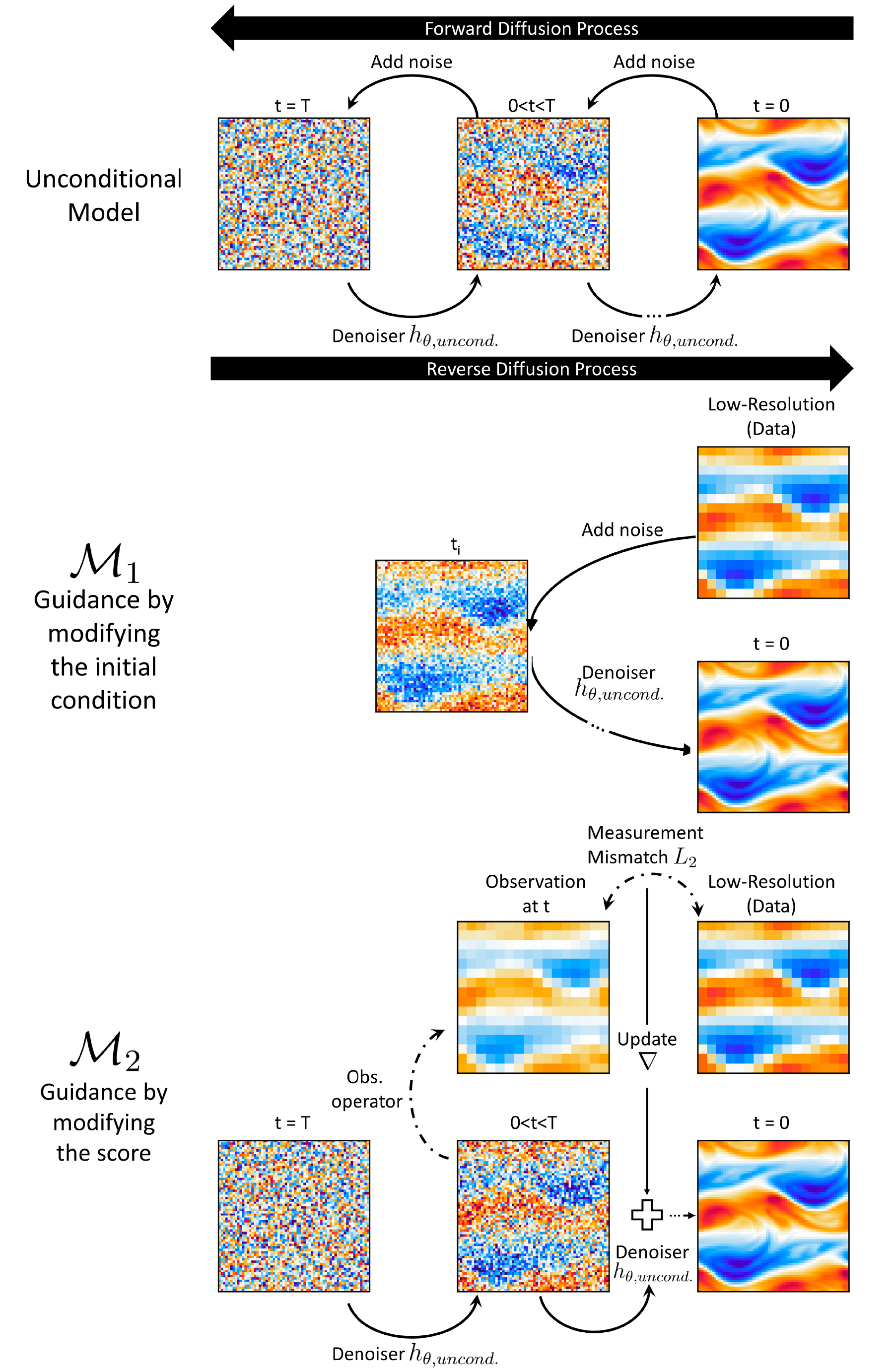} \\
    \caption{\small Schematic of the various unconditional diffusion modeling approaches. (Top row) Forward and reverse diffusion process for the unconditional model. (Middle row) Reverse process for guidance by modifying the initial condition (Sect.\;\ref{sec: SDEEdit}). (Bottom row) Reverse process for guidance by modifying the score (Sect.\;\ref{sec: DPS}).}
    \label{fig:schematic_uncond}
\end{figure}

One class of approaches \emph{guides} a pre-trained unconditional diffusion model to generate conditional samples. The main advantage is that these approaches can be directly applied to a pre-trained diffusion model and require no modification to the training process, and hence save significant computational resources. Fig.\;(\ref{fig:schematic_uncond}) shows a schematic of unconditional diffusion models and two ways to guide them.

\subsubsection{$\diffmodel{1}$: Modifying the Initial Condition (SDEdit)} \label{sec: SDEEdit}

In the first approach, $\diffmodel{1}$, the unconditional diffusion model is guided by simply starting the reverse diffusion process at an intermediate pseudo-time $\sdedittime$ and utilizing a new modified initial condition $\sdeditic$ that resembles noisy or corrupted samples generated at the intermediate pseudo-time $\sdedittime$.

As we will see, this approach, also known as \emph{SDEdit} \cite{meng2021sdedit}, directly preserves large-scale features of interest in the forward and reverse diffusion processes. The input low-resolution data $y \in \mathbb{R}^{\lrdim}$ is first interpolated to the target high-resolution to obtain $y_{interp} \in \mathbb{R}^{\hrdim}$. Next, samples from the forward process \eqref{eq:ito_forward} with the interpolated field $y_{interp}$ are obtained from the distribution \eqref{eq: forward process} at the intermediate pseudo-time $\sdedittime$, $0<\sdedittime<T$,
\begin{equation} \label{eq:sdedit}
     p_t(x_i) = \mathcal{N}(x_i;\mu_{\sdedittime} \, y_{interp},\sigma_{\sdedittime}^2 \mathcal{I}_{\hrdim}),
\end{equation}
This provides initial conditions $\sdeditic \in \mathbb{R}^{\hrdim}$ for the reverse diffusion.
Since this approach does not modify the score, the reverse unconditional diffusion process \eqref{eq:ito_reverse} is thus run with the unconditional score \eqref{eq: Tweedie unconditional} (learned using \eqref{eq: training_obj_uncond} and  \eqref{eq: training_score}) from this $\sdeditic$ at $\sdedittime$ back to the initial pseudo-time $t=0$. 
The final result is the conditional sampled high-resolution field $\hat{x}_0$. 
This approach is well-suited to super-resolution since the forward process adds noise to smooth out undesirable down-sampling artifacts while preserving large-scale flow features. 
A main parameter to be fine-tuned is the intermediate time $\sdedittime$, to ensure a sufficient trade-off between fidelity (sharpness) and cycle-consistency, i.e., do the filtered generated samples $\obsmodel(\hat{x}_0)$ match the low-resolution observation $y$? 

\subsubsection{$\diffmodel{2}$: Modifying the Score (Diffusion Posterior Sampling)} \label{sec: DPS}

In the second approach, the score of the unconditioned diffusion model is directly modified to generate conditional samples. The idea is to leverage Bayes' law
\begin{equation}
 p_t(x_t|y) =  \frac{p_t(y|x_t) p_t(x_t)} {p_t(y)}
\end{equation}
and hence derive
\begin{equation} \label{eq: score_DPS}
\nabla_{x_t}\bigg(\log p_t(x_t|y)\bigg)_{\diffmodel{2}} = \nabla_{x_t}\bigg(\log p_t(x_t)\bigg) + \nabla_{x_t}\bigg(\log p_t(y|x_t)\bigg)
\end{equation}
where the first term on the right side is simply the unconditional score \eqref{eq: Tweedie unconditional} that we have access to, and the second term is a \emph{measurement matching} term that guides the diffusion model to generate conditional samples that are consistent with the low-resolution input field (observations). However, this measurement matching term is analytically intractable and requires further simplification for practical implementations. \citeA{daras2024survey} provides a comprehensive review of approaches to approximate this term. One of the most popular approaches used is Diffusion Posterior Sampling (DPS) \cite{chung2023diffusion}, which assumes
\begin{equation} \label{eq: DPS_assumption}
\nabla_{x_t}\bigg(\log p_t(y|x_t)\bigg) \simeq \nabla_{x_t}\bigg(\log p_t(y|X_0=\mathbb{E}[X_0|X_t=x_t])\bigg)
\end{equation}
i.e., it assumes the score is the same if the conditioning is on the denoised samples instead of conditioning on the noisy or corrupted samples.
Since our measurement model \eqref{eq:prob_state} has a Gaussian structure, the likelihood $p_t(y|X_0=x_0)$ can be derived as
\begin{equation} \label{eq: likelihood}
  p_t(y|X_0=x_0) = \mathcal{N}(y;\obsmodel(x_0),\sigma_y^2 \mathcal{I}_{\lrdim})
\end{equation}
where $\obsmodel$ is the nonlinear observation operator and $\sigma_y$ models the noise in the observation process.
Hence, the likelihood $p_t(y|X_0=\mathbb{E}[X_0|X_t=x_t])$ can be derived from \eqref{eq: DPS_assumption} and \eqref{eq: likelihood} as
\begin{equation}
p_t(y|X_0=\mathbb{E}[X_0|X_t=x_t]) \approx \mathcal{N}(y;\obsmodel(\mathbb{E}[X_0|X_t=x_t]),\sigma_y^2 \mathcal{I}_\lrdim)
\end{equation}
\begin{equation} \label{eq: DPS_mm}
   \implies \nabla_{x_t}\bigg(\log p_t(y|x_t)\bigg) \simeq - \nabla_{x_t}\bigg(\frac{1}{2\sigma_y^2}{\norm{y - \obsmodel(\mathbb{E}[X_0|X_t=x_t])}}\bigg)
\end{equation}
The term ${y - \obsmodel(\mathbb{E}[X_0|X_t=x_t])}$ is the \emph{innovation vector} in data assimilation. Computing \eqref{eq: DPS_mm} requires access to gradients and Jacobian-vector products if $\obsmodel$ is nonlinear and if analytical closed-form expressions are inaccessible. Practically, computing \eqref{eq: DPS_mm} leads to instabilities at pseudo-times $t$ close to $0$ \cite{rozet2023score}. Hence, this measurement matching term is scaled by replacing the measurement noise $\sigma_y^2$ with  a more general scaled covariance approximation $\sigma_{y_{scaled}}^2$ that varies with the pseudo-time through a dependence on the noise schedule $\sigma_t$ to controllably guide the reverse diffusion process \cite{rozet2023score,manshausen2024generative},
\begin{equation}\label{eq: DPS_scaling}
    \sigma_{y_{scaled}}^2 =   \sigma_y^2 + \frac{\sigma_t^2}{\mu_t^2} C
\end{equation}
where $C$ is a tunable parameter that controls the guidance strength.
Hence, the final modified score is given by
\begin{equation} \label{eq: DPS_final}
    \nabla_{x_t}\bigg(\log p_t(x_t|y)\bigg)_{\diffmodel{2}} = \nabla_{x_t}\bigg(\log p_t(x_t)\bigg) -\frac{1}{2\sigma_{y_{scaled}}^2} \nabla_{x_t}\bigg({\norm{y - \obsmodel(\mathbb{E}[X_0|X_t=x_t])}}\bigg) 
\end{equation}
%
%
The first part of the right-hand side of \eqref{eq: DPS_final} can be learned using \eqref{eq: training_obj_uncond} and  \eqref{eq: training_score}, while the second part can be computed from \eqref{eq: DPS_mm} and \eqref{eq: DPS_scaling}. In summary, the second approach consists of running the reverse conditional diffusion process \eqref{eq:ito_reverse_conditional} with the score derived in \eqref{eq: DPS_final} starting from initial conditions $x_T \sim p_T$, i.e, samples from the tractable distribution, and the data $y$, to obtain conditional sampled high-resolution fields $\hat{x}_0$. 

\subsection{Conditional Models} \label{sec: conditional}

We now describe approaches that directly train conditional diffusion models using data pairs of the conditioning low-resolution field and target high-resolution field. 

\begin{figure}
        \includegraphics[width=1\linewidth]{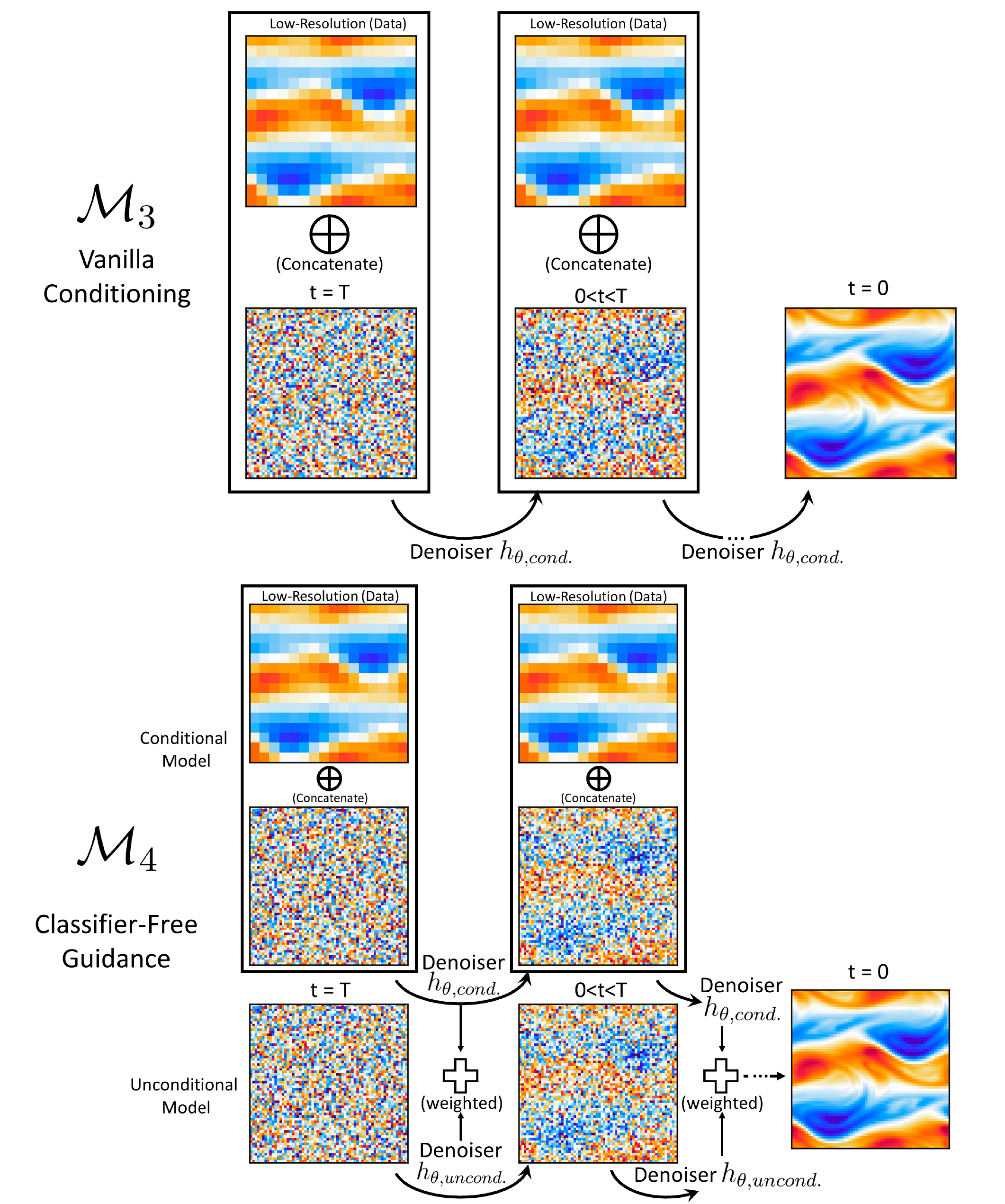} \\     
     \caption{\small Schematic of the various conditional diffusion modeling approaches (Sect.\;\ref{sec: conditional}). (Top row) Reverse diffusion process for the vanilla conditional model. (Bottom row) Reverse diffusion process for classifier-free guidance.}
    \label{fig: Schematic_cond}
\end{figure}

In the vanilla variant, $\diffmodel{3}$,  direct conditioning is typically achieved by modifying the denoiser neural network to take the low-resolution conditioning field as an additional input \eqref{eq: training_obj_cond}. Hence, this approach is also called \emph{image-to-image} diffusion modeling \cite{saharia2022palette}. During sampling, the reverse conditional SDE \eqref{eq:ito_reverse_conditional} can be run using the conditional score \eqref{eq: Tweedie conditional} (learned using the denoiser \eqref{eq: training_obj_cond} and \eqref{eq: training_score}), starting from initial conditions $x_T \sim p_T$, i.e, samples from the tractable distribution, and the data\;$y$, to obtain conditional sampled high-resolution fields $\hat{x}_0$. 

The vanilla conditional model can further be modified with a 
guidance strength
to further improve the quality of samples and controllably guide the diffusion model. This can be achieved using \emph{classifier-free guidance}  (CFG), $\diffmodel{4}$,  where the score is modified to be a linear combination of the conditional score and the unconditional score \cite{ho2021classifier}, similarly to relaxation schemes \cite{lermusiaux_2.29_notes},
\begin{equation} \label{eq: CFG}
\nabla_{x_t}\bigg(\log p_t(x_t|y)\bigg)_{\diffmodel{4}} = \nabla_{x_t}\bigg(\log p_t(x_t)\bigg) + w \bigg[\nabla_{x_t}\bigg(\log p_t(x_t|y)\bigg) - \nabla_{x_t}\bigg(\log p_t(x_t)\bigg) \bigg]
\end{equation}
where the parameter $w\geq1$ corresponds to the guidance strength. It determines how strongly the conditional model guides the reverse process. The denoiser for classifier-free guidance can be trained from scratch by simply training a vanilla conditional model \eqref{eq: training_obj_cond} with the conditioning being set randomly to a null field $\varnothing$ (which could be a field of zeros or pure Gaussian noise) every few instances of training to resemble an unconditional diffusion model. 
During sampling, the reverse SDE \eqref{eq:ito_reverse_conditional} can be run using the modified conditional score \eqref{eq: CFG} (learned using \eqref{eq: training_obj_uncond}, \eqref{eq: training_obj_cond}, and \eqref{eq: training_score}), starting from initial conditions $x_T \sim p_T$, i.e, samples from the tractable distribution, and the data $y$, to obtain conditional sampled high-resolution fields $\hat{x}_0$. 

Table\;(\ref{table:comp}) summarizes the differences among the four diffusion modeling approaches we consider and answers the questions we had asked. All methods except $\diffmodel{1}$ modify the unconditional score. In $\diffmodel{1}$, the initial conditions are simply modified to preserve large-scale features from the data (observations), making it more prone to failure than the other approaches. $\diffmodel{2}$ requires access to the observation operator and its gradients, which may not be possible when observations involve indirect or correlated variables, e.g., inference of ocean velocity or density fields from sea surface height observations \cite{souza2025surface}. Recently, methods such as BlindDPS \cite{chung2023parallel} perform posterior sampling from unknown measurement operators, though these methods may require a neural surrogate for the observation operator. The two conditional approaches require task-specific re-training, since the conditioning field varies based on the task (e.g., inference from a field obtained using a different observation operator). These models directly learn the complex observation-to-target map, and are particularly powerful when the conditioning fields are mutually informative fields obtained from highly nonlinear operators. All methods except $\diffmodel{3}$ contain tuning parameters to control guidance (Sect.\;\ref{sec: results-sensitivity}). In terms of cost, $\diffmodel{1}$, $\diffmodel{2}$ and $\diffmodel{3}$ require one neural network evaluation per pseudo-time step of the reverse process, while $\diffmodel{4}$ requires two evaluations. Since $\diffmodel{1}$ starts from an intermediate time, its cost scales with the number of pseudo-time steps, while $\diffmodel{2}$ incurs additional overhead from Jacobian and gradient computation.

\begin{table}[ht]
\setlength{\tabcolsep}{3pt} 
  \caption{Comparison of implementation and sampling characteristics of the four diffusion modeling approaches considered for super-resolution and inference. 
  }
  \scriptsize 
  \centering
  \begin{tabular}{l|cc|cc}
    \hline
     & \multicolumn{2}{c|}{\;Guided unconditional models}
       & \multicolumn{2}{c}{\;Conditional models } \\
    \cline{2-5}
     & \shortstack{$\diffmodel{1}$}
       & \shortstack{$\diffmodel{2}$}
       & $\diffmodel{3}$ & $\diffmodel{4}$ \\
    \hline
     Modifies the score                           & \xmark & \cmark & \cmark & \cmark \\
     Avoids need of observation operator   & \cmark & \xmark & \cmark & \cmark \\
      Task‐specific re‐training                    & \xmark & \xmark & \cmark & \cmark \\
     Controllable guidance               & \cmark & \cmark & \xmark & \cmark \\
     Number of function evaluations$^{a}$         & $\frac{\sdedittime}{T}$ & 1 & 1 & 2 \\
    \hline
    \multicolumn{5}{l}{$^{a}$ Relative number of denoiser evaluations per pseudo-timestep during the reverse diffusion process (sampling).}   \\
    \multicolumn{5}{l}{$\diffmodel{1}$: Modifying the initial condition \,(Sect.\;\ref{sec: SDEEdit}),\quad $\diffmodel{2}$: Modifying the score\,(Sect.\;\ref{sec: DPS})} \\
    \multicolumn{5}{l}{$\diffmodel{3}$: Vanilla conditional model\,(Sect.\;\ref{sec: conditional}),\quad 
                       $\diffmodel{4}$: Classifier‐free guidance\,(Sect.\;\ref{sec: conditional})}
  \end{tabular}
  \label{table:comp}
\normalsize
\end{table}

\section{Quasi-Geostrophic Turbulence and Data} \label{sec: Methods}

We now describe our high-resolution quasi-geostrophic turbulence simulations, including the dynamical regimes, numerical schemes, and parameters (Sects.\;\ref{sec: QG}--\ref{sec: numerics}). 
The measurement models we employ (Sect.\;\ref{sec: filtering}) then apply Large-Eddy Simulation (LES)-style spectral filtering and sparse observation operators to obtain low-resolution data (the observations).
In Sect.\;(\ref{sec: data-gen}), we describe the super-resolution and inference test cases and the high- and low-resolution data sets that are used to train the score-based diffusion models. In Sect.\;(\ref{sec: skill}), we introduce a comprehensive suite of skill-metrics used to assess the performance of the score-based diffusion models.
Once trained, these models will be used to reconstruct high-resolution fields from low-resolution data (Sect.\;\ref{sec: results}). 

\subsection{Quasi-Geostrophic Dynamics} \label{sec: QG}

As a testbed for our generative models, we use the two-dimensional (2D) incompressible one-layer Quasi-Geostrophic (QG) equations, an idealized approximation for large-scale geophysical flows on a rotating surface \cite{vallis2017atmospheric}. These partial differential equations (PDEs) are widely used to simulate atmospheric and oceanic flows, and have been utilized for multiple machine learning applications such as surrogate modeling \cite{li2020fourier} and subgrid-scale closure modeling \cite{frezat2022posteriori, ross2023benchmarking, guan2024online, srinivasan2024turbulence, suresh_babu_et_al_Oceans2025} under various approximations.

In 2D, the non-dimensionalized QG PDE can be written in vorticity form as
\begin{equation} \label{eq:QG}
    \frac{\partial \omega}{\partial t} + J(\psi,\omega) = \frac{1}{Re} \nabla^2 \omega - \mu \omega - \beta \frac{\partial \psi}{\partial x} + F
\end{equation}
where $\omega$ is the vorticity, which is our 2D field of interest, and $\psi$ is the stream function. In this governing PDE \eqref{eq:QG},
the nonlinear term $J(\psi, \omega)$ represents advection of vorticity by the flow itself. 
The diffusion term $\frac{1}{Re} \nabla^2 \omega$ represents turbulent eddy diffusivity, where $Re$ is the Reynolds number. 
The $\beta$-term accounts for latitudinal variation of the Coriolis force due to the Earth's rotation, while the linear damping term $\mu\omega$ models linear bottom drag \cite{stommel1948westward}. The forcing $F$ injects energy at prescribed length scales to mimic sustained geophysical driving. Since the flow is assumed to be incompressible, the velocity field $(u,v)$ can be recovered from the stream function via
\begin{eqnarray}
     & (u,v) = (-\frac{\partial \psi }{\partial y}, \frac{\partial \psi }{\partial x}) \\
        & \omega = \nabla^2 \psi
\end{eqnarray}
To drive the system and set dominant length scales, we apply a periodic forcing,
\begin{equation}  \label{eq:forcing}
    F = k_f[cos(k_f\;x) + cos(k_f\;y)]
\end{equation}
where $k_f$ is the forcing wavenumber \cite{guan2023learning}.
We consider all terms in \eqref{eq:QG} to simulate forced-dissipative turbulence and its 
vorticity dynamics.  

We emphasize two flow regimes: the \emph{eddy} regime and the \emph{jet} regime \cite{ross2023benchmarking}. The \emph{eddy} regime corresponds to $\beta=0$, where the flow is isotropic and comprised of large coherent vortices.
The \emph{jet} regime corresponds to $\beta > 0$, where the flow consists of alternatively banded jets. In the \emph{jet} regime, the $\beta$-effect forces the flow to organize into Rossby waves, creating a barrier to the upscale turbulent cascade of energy, leading to the destruction of coherent vortices \cite{rhines1975waves}. This interaction between Rossby waves and Quasi-Geostrophic turbulence is known as the Rhines effect, and has been observed in the Earth's oceans \cite{galperin2004ubiquitous} and planetary circulations of Jupiter and Saturn \cite{williams1978planetary, theiss2006generalized}.

\subsection{Numerical Schemes and Parameters} \label{sec: numerics}

We simulate 2D vorticity fields by solving \eqref{eq:QG}
on a square domain of dimensions $[0,2\pi] \times [0,2\pi]$ with periodic boundary conditions. We numerically solve \eqref{eq:QG} using a pseudo-spectral method \cite{orszag1974numerical}, with full $2/3$ de-aliasing. For numerical stability, we use a semi-implicit second-order Adams-Bashforth-Crank-Nicolson (AB2-CN) scheme for time-stepping, where the diffusion, Coriolis, and bottom drag terms are treated implicitly while the advection and forcing terms are treated explicitly \cite{boyd2001chebyshev}.

Following \citeA{graham2013framework} and \citeA{frezat2022posteriori}, we use the following scales for non-dimensionalization, corresponding to ocean mesoscales: a length-scale of $\frac{504}{\pi} \times 10^4$\;m and a time-scale of $1.2 \times 10^6$\;s. We simulate fully-resolved (FR) vorticity fields using a $N_{FR} \times N_{FR} = 512 \times 512$ numerical grid ($\Delta_{FR} = \frac{2\pi}{N_{FR}})$. To initialize simulations, we follow \citeA{guan2022stable} and sample initial vorticity fields from a standard normal distribution $\omega \sim \mathcal{N}(0,1)$ with wavenumbers $k \in [3,10]$, omitting small wavenumbers for numerical stability. We vary the Reynolds number between $Re=10^3$ and $Re=10^4$. Higher values of the Reynolds number increase the \emph{richness} of turbulence, leading to a wider range of multi-scale interactions \cite{pope2001turbulent}. For the \emph{eddy} regime, we set $\beta = 0$, and for the \emph{jet} regime, we set $\beta = 2.5$. 
For time-stepping, we use a non-dimensional time-step $dt$ a function of $Re$ and simulate a total time of $T = 100$.
A list of non-dimensional parameters is provided in Table\;(\ref{table: nondimparams}).   

\begin{table}[ht]
\setlength{\tabcolsep}{3pt}
\scriptsize
    \caption{Parameters of the different fully-resolved (FR) forced-dissipative Quasi-Geostrophic (QG) simulations for the eddy and jet regimes. All parameters are expressed in non-dimensional units defined in Sect.\;\ref{sec: numerics} (length-scale of $\frac{504}{\pi} \times 10^4$\;m and time-scale of $1.2 \times 10^6$\;s).}
    \centering
    \begin{tabular}{c c c c c c}
        \hline
        Regime & Re & $\beta$ & dt & $\mu$ & $k_f$ \\ \hline
        Eddy & $10^3$ & 0 &$1 \times 10^{-3}$ & $2 \times 10^{-2}$ & 2\\
        Eddy & $10^4$ & 0 & $5 \times 10^{-4}$ & $2 \times 10^{-2}$ & 2\\
        Jet & $10^3$ & 2.5 &$1 \times 10^{-3}$ & $2 \times 10^{-2}$ & 2\\
        Jet & $10^4$ & 2.5 &$5 \times 10^{-4}$ & $2 \times 10^{-2}$ & 2\\
        \hline
    \end{tabular}
    \label{table: nondimparams}
\normalsize
\end{table}

\subsection{Down-Sampling and Sparse Observation Operators} 
\label{sec: filtering}

For down-sampling, we apply coarsening operators on the vorticity fields obtained from the fully-resolved (FR) simulations. This coarsening operator could be of various types, and could act either in the physical space or the wavenumber space (Fourier domain). Down-sampling operations in physical space, such as max-pooling or average-pooling, have been considered in previous work \cite{fukami2019super}, but these correspond to non-local box filters in the wavenumber space, leading to spurious oscillations in the kinetic energy spectrum at high wavenumbers. In this work, we use spectral filtering operators utilized in Large-Eddy Simulations (LES) \cite{pope2001turbulent}. We apply our down-sampling operator ($\obsmodel_{\text{full}}$) in wavenumber space by first applying a Gaussian filter ($\obsmodel_{\text{Gaussian}}$), followed by a cut-off filter ($\obsmodel_{\text{cut-off}}$), and then interpolate the resultant field to obtain the observed field (OF) on the low-resolution grid with resolution $\Delta_{OF} = \frac{2\pi \times \delta}{N_{FR}}$:
\begin{eqnarray}\label{eq:downsample}
     \obsmodel_{\text{Gaussian}}(k_{FR}) =  \exp{\bigg(\frac{-k_{FR}^2\;(\delta \times \Delta_{FR})}{6}\bigg)} \\ \nonumber
    \obsmodel_{\text{cut-off}}(k_{FR}) = 0 \; , \qquad \forall\;k_{FR} > \frac{\pi}{\Delta_{OF}}  \\ \nonumber
    \obsmodel_{\text{full}} \triangleq  \text{interpolate}_{\Delta_{OF}}\circ \obsmodel_{\text{cut-off}}\circ\obsmodel_{\text{Gaussian}} 
\end{eqnarray} 
where $k_{FR}$ is the wavenumber corresponding to the fully-resolved grid and $\delta$ is the down-sampling scale (i.e., down-sampling from $64\times 64$ to $16\times16$ implies $\delta$ = $4$). The Gaussian filter smoothly attenuates high wavenumber content, resembling subgrid-scale energy removal, while the sharp cut-off filter enforces a strict wavenumber limit. This combination retains large-scale coherent structures while discarding small-scale features. Such filtering is crucial to pose a realistic and physically meaningful super-resolution problem for geophysical turbulence \cite{zhou2019subgrid, srinivasan2024turbulence}. 

We also study the additional effect of sparse and gappy observing systems ($\obsmodel_{\text{sparse}}$) by defining a set of unobserved regions $\Omega$ and using a mask operator in physical space,
\begin{eqnarray}\label{eq:downsample_coarse}
      \obsmodel_{\text{partial}}(x,y) = \omega_{fill} \; , \qquad \forall\;{x,y} \in \Omega \; \\ \nonumber
      \obsmodel_{\text{sparse}}(x,y) =  \obsmodel_{\text{partial}} \circ \obsmodel_{\text{full}}
\end{eqnarray}
where $\omega_{fill}$ is a fill value corresponding to the unobserved regions. In our case, we set the fill value to $0$, which is the mean of our vorticity fields over space and time. \citeA{martin2025generative} and \citeA{li2023multi} considered a similar partial observation operator to emulate satellite observations and study flow reconstruction with \emph{gappy} high-resolution data, respectively; however, we emphasize that here we apply these partial observations directly to the coarse-resolution output of the down-sampling operators defined in this section.

\subsection{Data and Test Cases}
\label{sec: data-gen}

To generate data for training, we simulate $500$ different trajectories of the fully-resolved (FR) QG simulations ($N_{FR} = 512$) with different random initial conditions by varying the random seed. Training diffusion models is very \emph{data-intensive}, since it requires sampling from the entire distribution of flow fields \cite{wang2023patch}. Hence, such large-dimensional flow fields require a lot of simulation data and computational resources for training. Therefore, we restrict our experiments to fields of dimensions $N_{FF} \times N_{FF} = 64 \times 64$, which are obtained by applying the down-sampling operator \eqref{eq:downsample} with $\delta = 2^3$ to these simulations. These new down-sampled fields, i.e, the \emph{filtered fields} (FF) serve as the high-resolution ground truth for our super-resolution experiments. This reduces the number of snapshots and hence computational costs (in terms of both memory and number of operations) required for training. Alternatively, the size of the high-resolution target can also be restricted by simply selecting fields from a smaller window of the original fully-resolved fields. However, the size of this window needs to be larger than the largest eddy of the simulated vorticity field \cite{fukami2019super}.

Snapshots (frames) of the vorticity field are saved every $\Delta T_{save} = 0.5$ non-dimensional time units, resulting in $\frac{T}{\Delta T_{save}} = 200$ snapshots. Upon analyzing the energy spectrum, we observe self-similarity after $t = 50$, and only utilize snapshots obtained post this time for our training and analysis (last 100 snapshots). Hence, our generative models are trained with $500 \times 100 = 50,000$ snapshots. 

\begin{figure}
    \centering
    \includegraphics[width=1\linewidth]{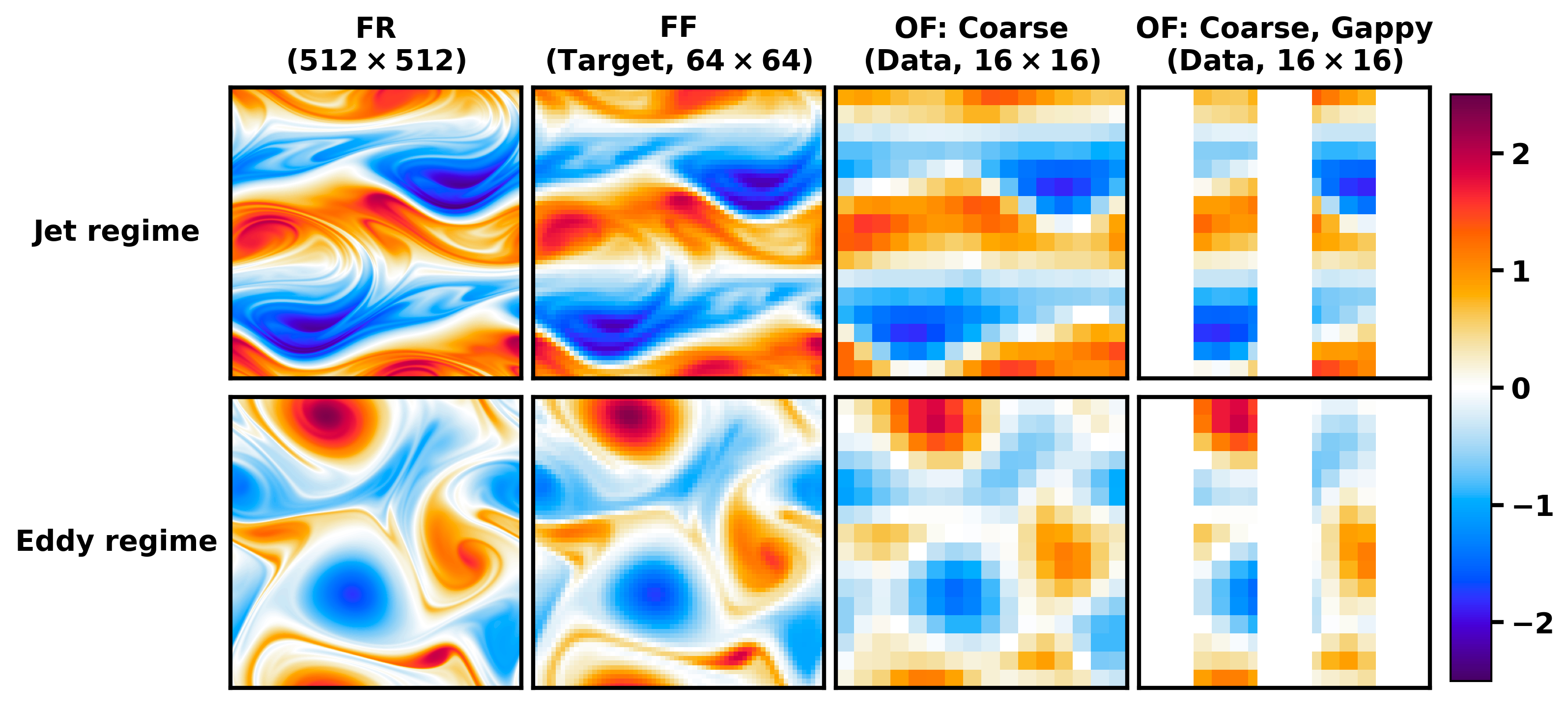}
    \caption{\small Test case description for super-resolution and inference in the jet (Top row) and eddy (Bottom row) regimes at $Re=10^4$. Column 1: Fully-resolved (FR) vorticity field. Column 2: Filtered-field (FF), which serves as the target for super-resolution. Column 3: Observed field (OF) under coarse-resolution observations. Column 4: Observed field (OF) under coarse, sparse and gappy observations. The observed fields serve as data (input) for super-resolution.}
    \label{fig: test_cases}
\end{figure}

To study the performance of generative models 
(Table\;\ref{table:comp}), we utilize test cases with two observation operators for all the simulations in Table\;(\ref{table: nondimparams}): down-scaling by a scale of $\delta = 2^2$ with full observations \eqref{eq:downsample}, and a similar down-scaling with partial observations \eqref{eq:downsample_coarse}. To study the effect of sparse, gappy observations, we restrict the observed region $\Omega$ to cover $40\%$ of the total domain. Utilizing the measurement model\;\eqref{eq:prob_state} with small Gaussian observation noise $\sigma_y = 0.01$, these test cases yield observed fields (OF) with dimensions $N_{OF} \times N_{OF} = 16 \times 16$, which serve as the conditioning field (input) to our diffusion models. Table\;(\ref{table: testcases}) summarizes the 8 test cases considered. We perform super-resolution at a fixed time, utilizing observed fields obtained from the $64 \times 64$ filtered vorticity fields of a trajectory initialized with a new initial condition, which was not utilized for training. A visual representation of test cases 2, 4, 6, and 8 is shown in Fig.\;(\ref{fig: test_cases}).

\begin{table}[ht]
\setlength{\tabcolsep}{3pt}
\scriptsize
    \caption{\small Main test cases utilized for super-resolution and inference with diffusion models corresponding to target dimensions ($N_{FF}=64 \times 64$) and observed field dimensions ($N_{OF} = 16 \times 16$), with resultant down-sampling scale ($\delta = 2^2$). The effect of other down-sampling scales ($\delta \in 2^1, 2^3$) are discussed in Sect.\;(\ref{sec: results-full}). 
    }
    \centering
    \begin{tabular}{c c c c}
        \hline
        Number & Regime & Re & Data (Observations) \\ \hline
        1 & Jet & $10^3$ & Coarse-resolution field  \\
        2 & Jet & $10^4$ & Coarse-resolution field   \\
        3 & Eddy & $10^3$ & Coarse-resolution field   \\
        4 & Eddy & $10^4$ & Coarse-resolution field   \\
        5 & Jet & $10^3$ &  Coarse and sparse, gappy  \\
        6 & Jet & $10^4$ & Coarse and sparse, gappy  \\
        7 & Eddy & $10^3$ & Coarse and sparse, gappy   \\
        8 & Eddy & $10^4$ & Coarse and sparse, gappy   \\     
        \hline   \\
    \end{tabular}
    \label{table: testcases}
\normalsize
\end{table}

\subsection{Skill Metrics} \label{sec: skill}

To assess the performance of the score-based diffusion models, we utilize three sets of complementary skill metrics: one is related to the generated vorticity fields, the second to statistical turbulence quantities, and the third to uncertainty quantification.

First, we assess the generated super-resolved fields using common norms such as the relative $L_2$ norm: $\frac{\normsqrt{\omega_{FF}-\hat{\omega}_0}}{\normsqrt{\omega_{FF}}}$ and measures of the accuracy of features of the reconstructed vorticity field (e.g., filament sharpness). We then check for cycle-consistency using the relative norm: $\frac{\normsqrt{\omega_{OF}-\obsmodel(\hat{\omega}_0)}}{\normsqrt{\omega_{OF}}}$, i.e, does the filtered high-resolution diffusion output match the low-resolution input field? Cycle-consistency is essential to ensure that the observed field can indeed be obtained from the generated super-resolved field through the measurement model  \cite{zhou2016learning}. 

Second, we compare time-averaged statistical quantities relevant to 2D turbulence. Following \citeA{frezat2022posteriori}, we compute the angle-averaged kinetic energy spectrum $E(k)$ and enstrophy spectrum $Z(k)$  in the wavenumber space using
\begin{equation} \label{eq: KE_spectra}
    E(k) \;=\;
    \frac{1}{2}
    \int_{|\mathbf{k}|=k}
    \hat{\mathbf{u}}_i(\mathbf{k})\; \hat{\mathbf{u}}_i^{*}(\mathbf{k})
    \;\mathrm{d}S(\mathbf{k})
\end{equation}
\begin{equation} \label{eq: Ens_spectra}
    Z(k) \;=\; 
    \frac{1}{2}
    \int_{|\mathbf{k}|=k}
    \hat\omega(\mathbf{k})\;\hat\omega^{*}(\mathbf{k})
    \;\mathrm{d}S(\mathbf{k})
\end{equation}
where  $k = \sqrt{k_x^2 + k_y^2}$ denotes the horizontal angular wavenumber, $\widehat{ \bullet}$ the Fourier transform, and $\bullet^{*}$ the complex conjugate. We also compute the relative $L_2$ norms of the log-spectra. Because the $\beta$-effect induces anisotropic spectral energy transfers that are obscured by angle-averaging  \cite{huang2001anisotropic}, we then compute contours of the pre-multiplied 2D kinetic energy spectrum,
\begin{equation} \label{eq: 2D_KE_spectra}
    E(k_x,k_y) \;=\;
    \lambda_x \; \lambda_y \; \frac{1}{2}
    \hat{\mathbf{u}}_i(\mathbf{k_x,k_y})\; \hat{\mathbf{u}}_i^{*}(\mathbf{k_x,k_y})
\end{equation}
where $\lambda_x = \frac{2\pi}{k_x}$ and $\lambda_y = \frac{2\pi}{k_y}$ are wavelength
factors to account for the logarithmic spacing in the wavenumber space \cite{towne2020resolvent}. We also evaluate the mean effective resolution using the time-averaged noise-to-signal ratio of the spectrum \cite{ballarotta2019resolutions},
\begin{equation} \label{eq: NSR_ratio}
    NSR(k) \;=\;  \frac{Z_{error}(k)} {Z_{FF}(k)}
\end{equation}
where $Z_{error}$ is is the enstrophy spectrum \eqref{eq: Ens_spectra} of the error, $\hat{\omega}_0 - {\omega}_{FF}$, and $Z_{FF}$ is that of the target vorticity, ${\omega}_{FF}$.
To study the non-Gaussian single-point statistics of turbulent flows, which is a statistical measure of intermittency \cite{mcwilliams2007extreme,wilczek2009dynamical}, we compute the one-dimensional probability distribution function for the vorticity field using kernel density estimation \cite{wkeglarczyk2018kernel}.

Third, as our diffusion models can generate an ensemble of super-resolved vorticity fields, they can be utilized for uncertainty quantification \cite{leinonen2023latent, chen2025taming}. To quantify and verify the uncertainty, we compute the ensemble mean, $\overline{\hat{\omega}_0}$ = $\frac{1}{N_{ens}}\sum\limits_{i=1}^{N_{ens}}\hat{\omega}_0^i$, ensemble standard deviation: $\sqrt{\frac{\sum\limits_{i=1}^{N_{ens}}(\overline{\hat{\omega}_0}-\hat{\omega}_0^i)^2}{N_{ens}}}$, and root-mean squared error (RMSE) of the generated super-resolved fields: $\sqrt{\frac{\sum\limits_{i=1}^{N_{ens}}\normsqrt{{{\omega}_{FF}}-\hat{\omega}_0^i}^2}{N_{ens}}}$, where $N_{ens} = 16$ is the ensemble size and the superscript $i$ indicates the $i^{\text{th}}$ ensemble member.

\section{Applications}  \label{sec: results}

To train the diffusion models, we utilize a U-Net architecture \cite{ronneberger2015u}. To provide conditioning, we concatenate the interpolated low-resolution field with the noisy field at each diffusion time-step following \citeA{saharia2022image}. Details of neural architectures, training, and associated hyperparameters are in \ref{sec: training}. 

Next, we present the results of the four generative diffusion models (Sect.\;\ref{sec: diffusion}) for super-resolution of turbulent vorticity fields from coarse-resolution fields (Sect.\;\ref{sec: results-full}) and then from coarse, sparse, and gappy observations (Sect.\;\ref{sec: results-sparse}). All these results are obtained using best-tuned intermediate times and guidance strengths. In Sect.\;(\ref{sec: results-sensitivity}), we study the sensitivity of the four diffusion models to their main tuning parameters.

\subsection{Applications 1: Super-resolution from Coarse-resolution Fields} \label{sec: results-full}

The first applications employ coarse-resolution vorticity fields as the observed field (data or input), corresponding to Cases 1-4 of Table\;(\ref{table: testcases}), and assess whether the diffusion models can reconstruct subgrid-scale structure from coarse data \cite{fukami2019super}.

In the jet regime (Cases 1-2 of Table\;(\ref{table: testcases})), strong $\beta$-effects organize the flows into anisotropic zonal bands with thin filaments, while in the eddy regime (Cases 3-4 of Table\;(\ref{table: testcases})), the absence of $\beta$-effects allows 
more complex, nearly isotropic dynamics with thin filaments and streaks surrounding
larger vortex cores
\cite{maltrud1991energy, kevlahan1997vorticity}. Table\;(\ref{table: metrics_jet_full}) summarizes the key quantitative skill metrics: relative error norms of the reconstructed fields, cycle-consistency error, and log-spectra, as well as the ensemble standard deviation of the reconstructed field for the jet and eddy regimes. On average, all models perform better in the jet regime; the presence of zonal bands helps the models identify the larger-scale structures than in the isotropic eddy regime.

\begin{table}[h!]
\setlength{\tabcolsep}{3pt}
\scriptsize 
\centering
\caption{\small Comparison of the key quantitative skill metrics for super-resolution from coarse-resolution fields (Cases 1-4 of Table\;(\ref{table: testcases})).} \hspace*{-0.85in}
\begin{tabular}{l|l|cccc|cccc}
  \toprule
  \multicolumn{2}{l|}{} 
  & \multicolumn{4}{c|}{$Re=10^3$}
  & \multicolumn{4}{c}{$Re=10^4$} \\
  \midrule
  Relative Error Metric
  & Scale
  & \shortstack{$\diffmodel{1}$}
    & \shortstack{$\diffmodel{2}$}
    & \shortstack{$\diffmodel{3}$}
    & \shortstack{$\diffmodel{4}$}
    & \shortstack{$\diffmodel{1}$}
    & \shortstack{$\diffmodel{2}$}
    & \shortstack{$\diffmodel{3}$}
    & \shortstack{$\diffmodel{4}$} \\
  \toprule
 \multicolumn{2}{c}{\hspace{-1in} Jet regime} & \multicolumn{4}{c}{} & \multicolumn{4}{c}{} \\
  \toprule
  Reconstructed vorticity ($\downarrow$)  & ($\times 10^{-1}$) 
    & 1.05 $\pm$ 0.08 & 0.45 $\pm$ 0.07 & \textbf{0.19 $\pm$ 0.01} & 0.20 $\pm$ 0.01
    & 2.13 $\pm$ 0.09 & 1.29 $\pm$ 0.07 & \textbf{0.69 $\pm$ 0.05} & 0.71 $\pm$ 0.02 \\
   Ensemble Field Std. & ($\times 10^{0}$) 
    & 0.05 & 0.04 & 0.01 & 0.01
    & 0.09 & 0.09 & 0.05 & 0.05 \\
 Cycle-consistency ($\downarrow$) & ($\times 10^{-1}$) 
    & 0.77 $\pm$ 0.05 & 0.30 $\pm$ 0.05 & \textbf{0.04 $\pm$ 0.01} & 0.05 $\pm$ 0.01
    & 1.31 $\pm$ 0.07 & 0.61 $\pm$ 0.06 & 0.05 $\pm$ 0.03 & \textbf{0.04 $\pm$ 0.00} \\
  Log-energy spectrum ($\downarrow$) & ($\times 10^{-2}$) 
    & 11.09 $\pm$ 1.16 & 4.14 $\pm$ 0.15 & \textbf{3.41 $\pm$ 0.40} & 3.67 $\pm$ 0.25
    & 7.29 $\pm$ 0.74 & \textbf{1.81 $\pm$ 0.18} & 2.11 $\pm$ 0.93 & 2.36 $\pm$ 0.37 \\
  Log-enstrophy spectrum ($\downarrow$) & ($\times 10^{-2}$) 
    & 17.09 $\pm$ 1.78 & 6.38 $\pm$ 0.23 & \textbf{5.25 $\pm$ 0.61} & 5.65 $\pm$ 0.38
    & 11.85 $\pm$ 1.21 & \textbf{2.95 $\pm$ 0.29} & 3.43 $\pm$ 1.51 & 3.85 $\pm$ 0.61 \\
  \toprule
  \multicolumn{2}{c}{\hspace{-1in} Eddy regime} & \multicolumn{4}{c}{} & \multicolumn{4}{c}{} \\
    \toprule
  Reconstructed vorticity ($\downarrow$)  & ($\times 10^{-1}$) 
    & 3.02 $\pm$ 0.08 & 2.65 $\pm$ 0.13 & \textbf{1.00 $\pm$ 0.06} & 1.01 $\pm$ 0.04
    & 3.40 $\pm$ 0.10 & 3.22 $\pm$ 0.11 & \textbf{1.52 $\pm$ 0.06} & 1.53 $\pm$ 0.07 \\
  Ensemble Field Std. & ($\times 10^{0}$) 
    & 0.12 & 0.17 & 0.06 & 0.06
    & 0.16 & 0.22 & 0.10 & 0.01 \\
 Cycle-consistency ($\downarrow$) & ($\times 10^{-1}$) 
    & 1.94 $\pm$ 0.07 & 1.39 $\pm$ 0.10 & 0.09 $\pm$ 0.06 & \textbf{0.06 $\pm$ 0.00}
    & 1.96 $\pm$ 0.10 & 1.42 $\pm$ 0.09 & 0.04 $\pm$ 0.01 & \textbf{0.03 $\pm$ 0.00} \\
  Log-energy spectrum ($\downarrow$) & ($\times 10^{-2}$) 
    & 6.31 $\pm$ 1.37 & \textbf{5.83 $\pm$ 0.20} & 6.27 $\pm$ 1.80 & 5.95 $\pm$ 0.94
    & 5.67 $\pm$ 0.37 & 2.79 $\pm$ 0.74 & 2.63 $\pm$ 0.32 & \textbf{2.51 $\pm$ 0.30} \\
  Log-enstrophy spectrum ($\downarrow$) & ($\times 10^{-2}$) 
    & 9.71 $\pm$ 2.12 & \textbf{8.99 $\pm$ 0.30} & 9.67 $\pm$ 2.77 & 9.18 $\pm$ 1.46
    & 9.20 $\pm$ 0.60 & 4.53 $\pm$ 1.20 & 4.26 $\pm$ 0.51 & \textbf{4.07 $\pm$ 0.48} \\
  \toprule
  \multicolumn{10}{l}{Error metrics are shown as ensemble mean $\pm$ one standard deviation; the best-performing model for each metric is highlighted in \textbf{bold}.}
\end{tabular}\label{table: metrics_jet_full}
\normalsize 
\end{table}

\emph{Super-resolved vorticity fields}: In both the jet and eddy regimes, and for both $Re=10^3$ and $Re=10^4$, the vanilla conditional model ($\diffmodel{3}$) and the classifier-free guidance approach ($\diffmodel{4}$) achieve the lowest reconstruction error, followed by the guided unconditional models: guidance by modifying the score (DPS, $\diffmodel{2}$) and guidance by modifying the initial conditions (SDEdit, $\diffmodel{1}$). All four methods capture the larger-scale forced features (zonal jets and vortex cores for the jet and eddy regimes, respectively) of the target field driven by the forcing in eqs.\;\eqref{eq:QG} --\eqref{eq:forcing}. 

\begin{figure}[] \hspace*{-0.75in}
    \includegraphics[width=1.2\linewidth]{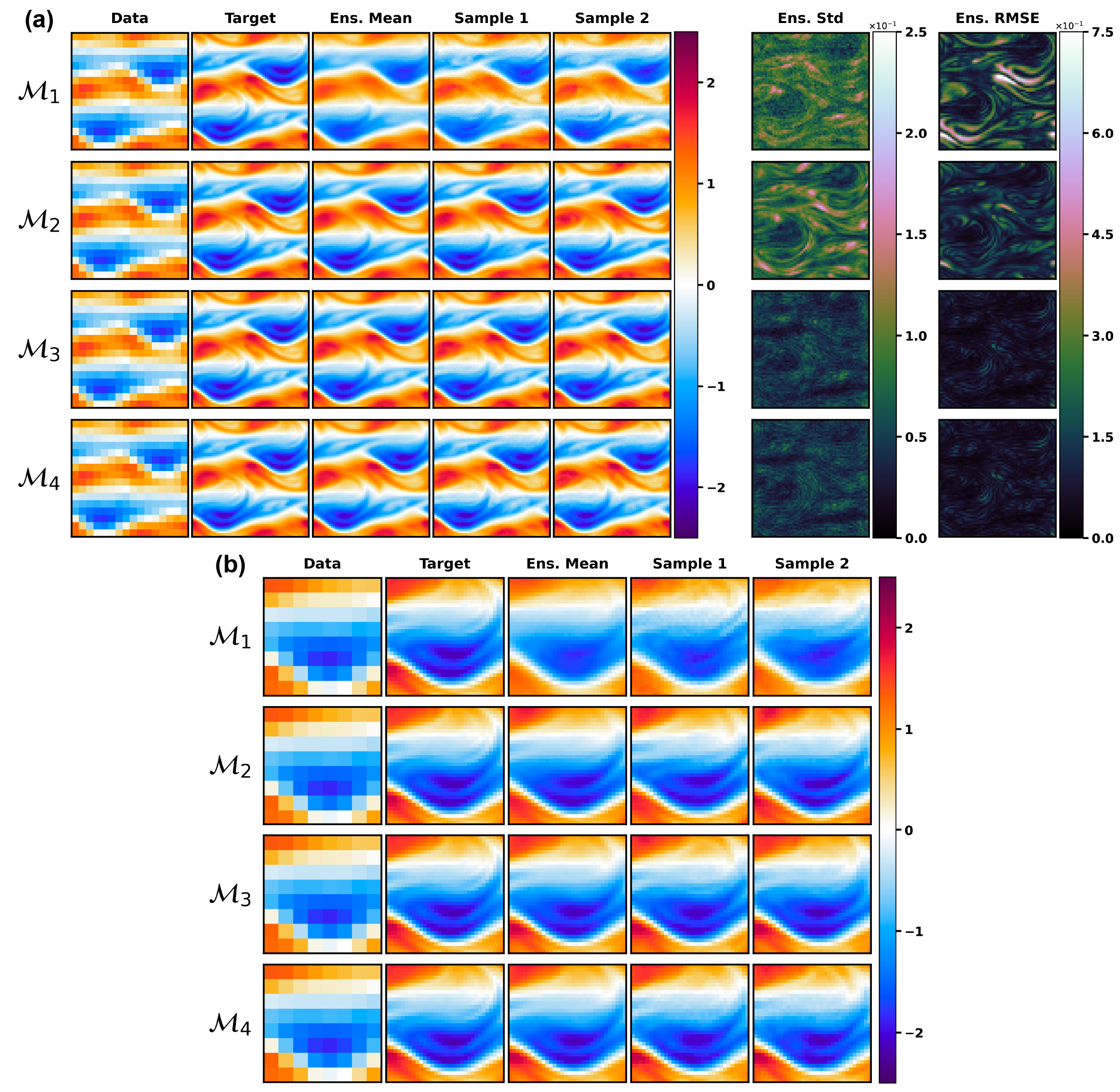}
    \caption{\small (a) Reconstructed vorticity fields for the four super-resolution approaches in the jet regime (Re=$10^4$) from coarse-resolution fields for one snapshot (Test Case 2, Table\;\ref{table: testcases}). Column\;1: Coarse observed vorticity field (data, OF). Column\;2: Corresponding filtered vorticity field (target, FF). Column\;3: Ensemble mean of the super-resolved diffusion model outputs. Columns\;4 and 5: Two representative ensemble members.
    Columns\;6 and 7: Ensemble point-wise standard deviation (Std.) and root-mean squared error (RMSE), respectively. (b) Zoomed-in view of the upper-right quadrant for the first 5 columns of (a), highlighting fine-scale features. Data and target are identical for all models.}
    \label{fig: fields_jet_full}
\end{figure}

\begin{figure}[] \hspace*{-0.75in}
        \includegraphics[width=1.2\linewidth]{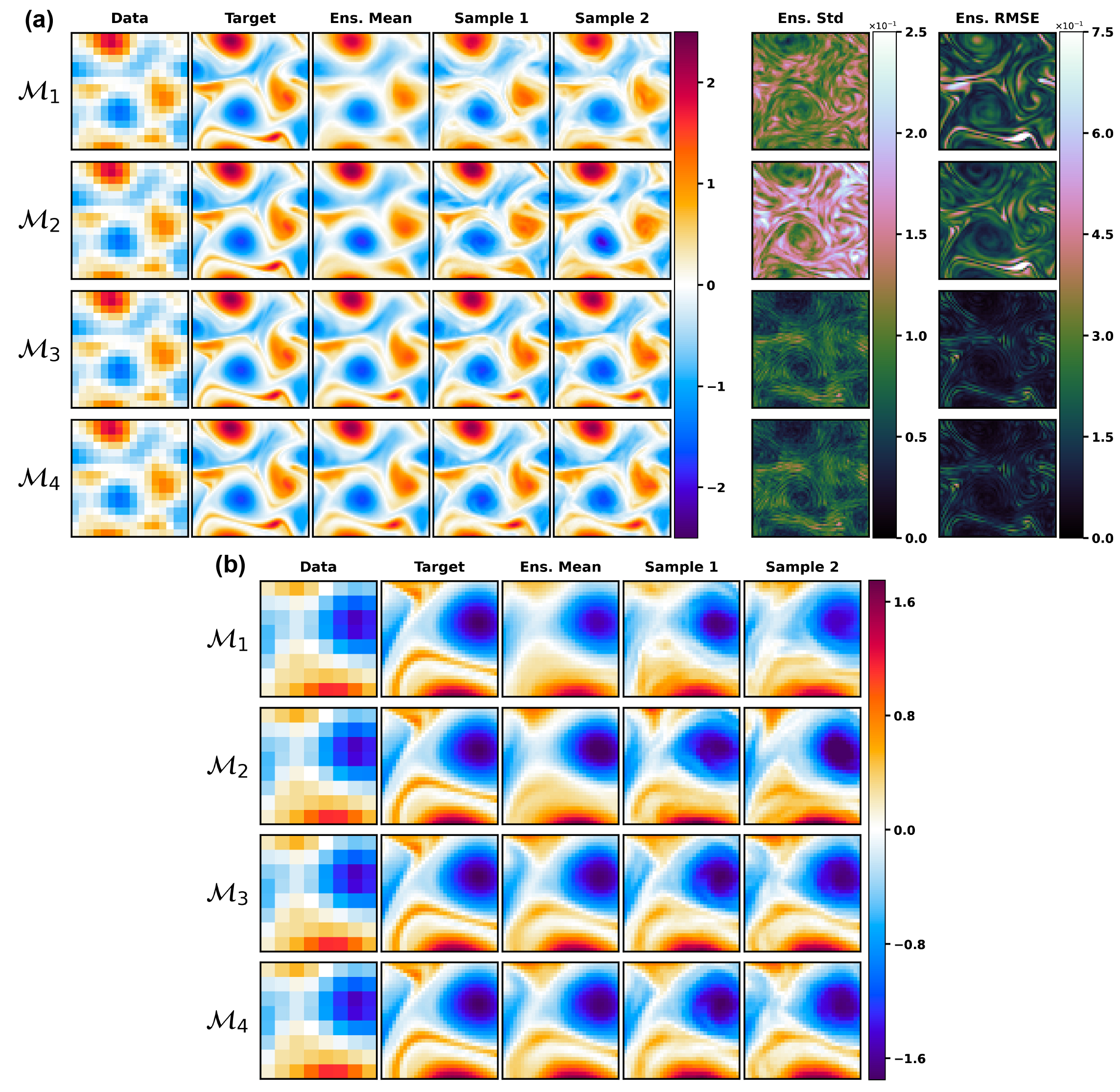}
   \caption{\small (a) As Fig.\;(\ref{fig: fields_jet_full} a), but for super-resolution in the eddy regime (Re=$10^4$) from coarse-resolution fields (Test Case 4, Table\;\ref{table: testcases}). (b) As Fig.\;(\ref{fig: fields_jet_full} b), zoomed-in view of the lower-left quadrant for the first 5 columns of (a), highlighting fine-scale features. Data and target are identical for all models.}
\label{fig: fields_eddy_full} 
\end{figure}

Figs.\;(\ref{fig: fields_jet_full} a) and (\ref{fig: fields_eddy_full} a) and their zoomed-in counterparts Figs.\;(\ref{fig: fields_jet_full} b) and (\ref{fig: fields_eddy_full} b) show that $\diffmodel{1}$ leads to overly smooth ensembles which do not capture boundaries (large vorticity gradient regions), or fine structure (zonal filaments and filaments with streaks in the jet and eddy regimes, respectively), because $\diffmodel{1}$ does not sample from the conditional reverse process \eqref{eq:ito_reverse_conditional}, and only uses large-scale data features \eqref{eq:sdedit}.
$\diffmodel{2}$'s ensemble better captures zonal jet and vortex core boundaries, but its mean still lacks fine structure (filaments at core edges) since $\diffmodel{2}$ only approximates the conditional distribution \eqref{eq: DPS_final} via assumption \eqref{eq: DPS_assumption}. Sampling directly from the conditional reverse process \eqref{eq:ito_reverse_conditional}, the conditional models $\diffmodel{3}$ and $\diffmodel{4}$ generate super-resolved ensembles qualitatively similar to the target that capture boundaries and most fine structures.


\emph{Uncertainty quantification}: 
From the last two columns of Figs.\;(\ref{fig: fields_jet_full} a) and (\ref{fig: fields_eddy_full} a), $\diffmodel{3}$ and $\diffmodel{4}$ exhibit the smallest standard deviation and RMSE (with $N_{ens} = 16$) versus the two guided unconditional models. 
For all models, standard deviation and error are considerably higher in the eddy regime than in the jet regime (Table\;\ref{table: metrics_jet_full}), due to the more complex distribution of isotropic vorticity fields.
Uncertainty is largest at the zonal jet and vortex core boundaries, where gradients are largest, since precise gradient super-resolution is nontrivial. Moreover, the standard deviations and errors show similar spatial patterns (except $\diffmodel{1}$), so these models can serve as probabilistic estimators, providing both accurate mean predictions and spatially consistent uncertainty quantification.
Similar spatial structures for vanilla conditional models were observed by \citeA{souza2025surface}.

\begin{figure}[htb]\hspace*{-1in}
        \includegraphics[width=1.4\linewidth]{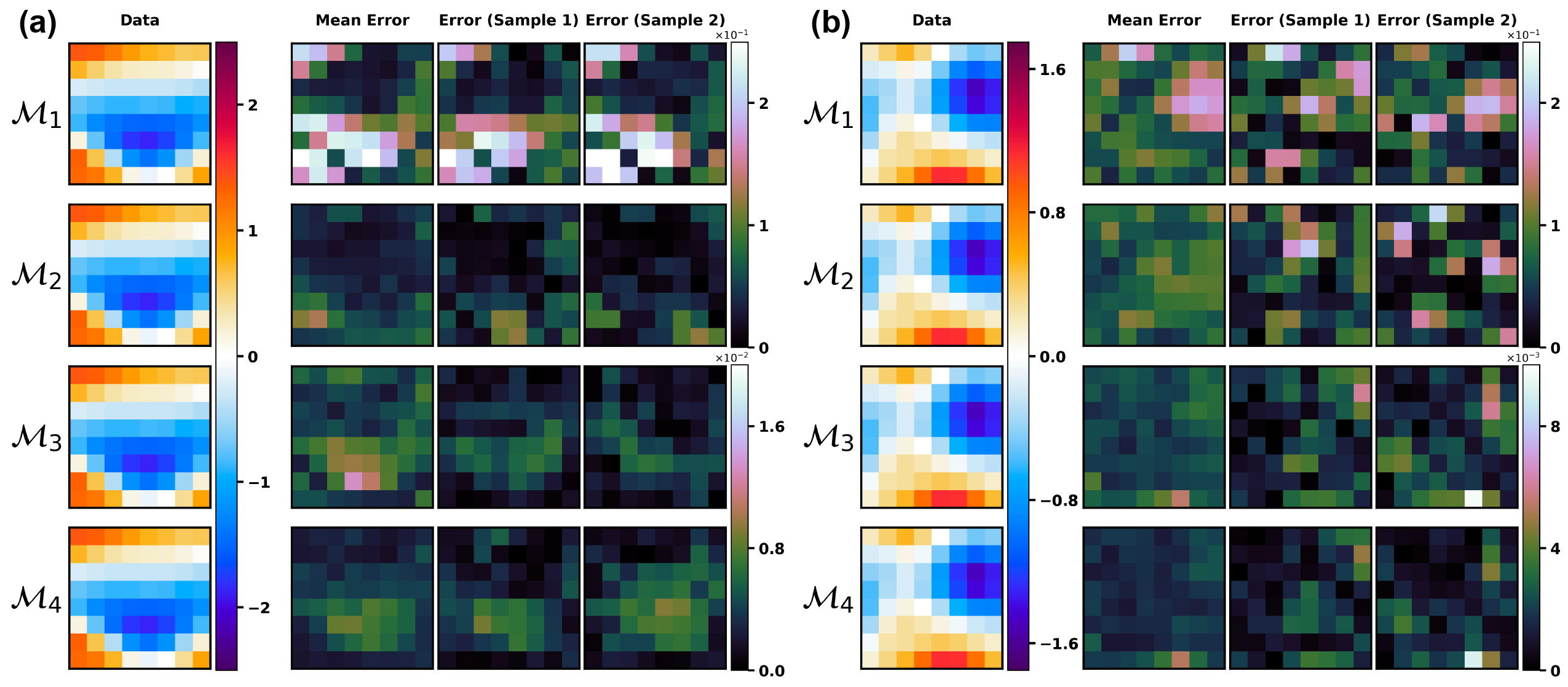} 
    \caption{\small Cycle-consistency of the four diffusion modeling approaches, starting from coarse-resolution fields (Re=$10^4$). Zoomed-in views of the (a) Upper-right quadrant for super-resolution in the jet regime (Test Case 2, Table\;\ref{table: testcases}) and (b) Lower-left quadrant for super-resolution in the eddy regime (Test Case 4, Table\;\ref{table: testcases}). Column 1: Coarse observed vorticity field (Data). Column 2: Ensemble mean of the errors of the filtered super-resolved diffusion output. Columns 3 and 4: Errors of 2 filtered ensemble members. The upper color bar limit of the error fields for $\diffmodel{1}$ and $\diffmodel{2}$ differs from that for $\diffmodel{3}$ and $\diffmodel{4}$ by orders of magnitude.
    }\label{fig: cons_jet_full}
\end{figure}

\emph{Cycle-consistency}: In both the jet (Fig.\;\ref{fig: cons_jet_full} a)  and eddy (Fig.\;\ref{fig: cons_jet_full} b) regimes, $\diffmodel{1}$ has region-dependent cycle-inconsistency as it samples from the unconditional \eqref{eq:ito_reverse}, and not the conditional \eqref{eq:ito_reverse_conditional}, reverse diffusion process. $\diffmodel{2}$ is less inconsistent than $\diffmodel{1}$ since it approximates the conditional distribution through the measurement-matching term \eqref{eq: DPS_mm}. Unfortunately, $\diffmodel{2}$ assumes that the conditional expectation from the noisy and denoised samples is equivalent \eqref{eq: DPS_assumption}, which fails near sharp image gradients generated towards the end of the reverse process and breaks cycle-consistency. The error thus remains one to two orders of magnitude larger than the conditional models (Table\;\ref{table: metrics_jet_full}).
Despite a lack of explicit guidance, the two conditional models $\diffmodel{3}$ and $\diffmodel{4}$ are extremely cycle-consistent, as they sample directly from the true conditional distribution, \eqref{eq:ito_reverse_conditional} and \eqref{eq: CFG}, respectively. While tuning the guidance strength of $\diffmodel{4}$ could also lead to sampling from unphysical distributions \cite{karras2024guiding}, additional conditioning helps avoid some artifacts seen with $\diffmodel{3}$.

\begin{figure}[]\hspace*{-1in}
    \includegraphics[width=1.4\linewidth]{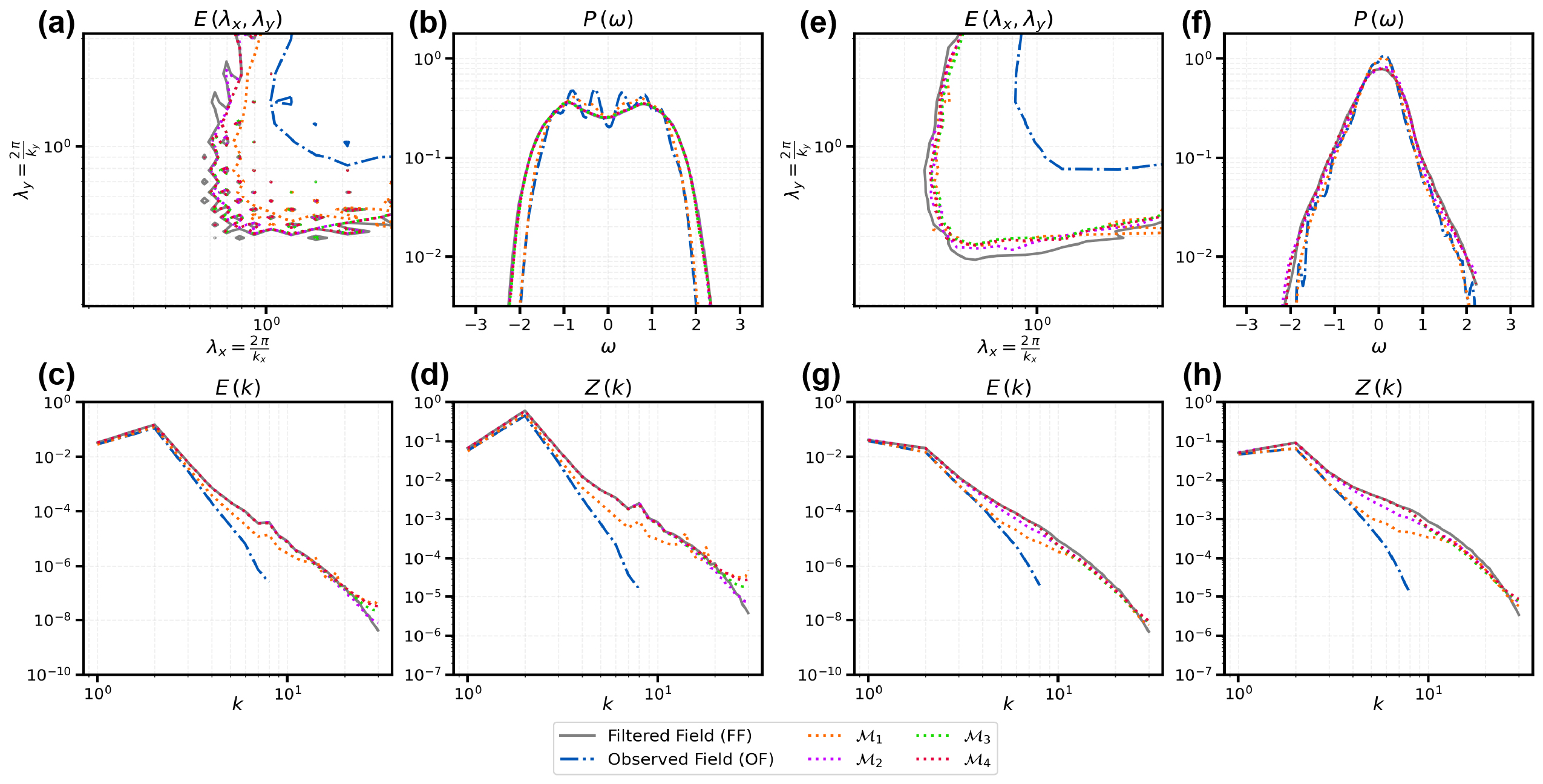}
     \caption{\small Time-averaged statistical quantities of the four models for super-resolution from coarse-resolution fields in the (a-d) jet regime (Re=$10^4$) (Test Case 2, Table\;\ref{table: testcases}) and (e-h) eddy regime (Re=$10^4$) (Test Case 4, Table\;\ref{table: testcases}). (a, e): Premultiplied 2D kinetic energy spectrum contours that capture $95\%$ of the energy. (b, f): Single-point probability distribution function for the vorticity field. (c, d): Angle-averaged kinetic energy and (g, h) enstrophy spectra. Solid lines correspond to the filtered field (target), dash-dot lines to the observed field (data), and dotted lines to the diffusion model outputs. }
    \label{fig: spectra_jet_full}
\end{figure}

\emph{Time-averaged 2D turbulence statistics}: The single-point vorticity distribution of both the jet (Fig.\;\ref{fig: spectra_jet_full} b) and eddy (Fig.\;\ref{fig: spectra_jet_full} f) regimes shows non-Gaussianity, arising from the intermittency of long-lived coherent vortices in two-dimensional turbulence \cite{arbic2003coherent, srinivasan2024turbulence}. 
The unimodal eddy regime vorticity distribution becomes bimodal in the jet regime due to the zonal jets, examples of mixing barriers that introduce multimodality \cite{david2017statistical}. $\diffmodel{2}$, $\diffmodel{3}$, and $\diffmodel{4}$ accurately capture the single-point distributions in both regimes. The distribution tails are associated with how well the underlying coherent structures, such as vortices, are captured \cite{farge1999non, mcwilliams2007extreme}. $\diffmodel{1}$ is skewed towards the noisy observations used for initialization, and underestimates these rare events, as in Figs.\;(\ref{fig: fields_jet_full}) and (\ref{fig: fields_eddy_full}). 

Figs.\;(\ref{fig: spectra_jet_full} a) and (\ref{fig: spectra_jet_full} e) show the premultiplied 2D kinetic energy spectrum, with its characteristic jagged and smooth contours for the jet and eddy regimes, respectively \cite{nozawa1997spectral, huang2001anisotropic}. $\diffmodel{2}, \diffmodel{3}$ and $\diffmodel{4}$ capture the spectrum shape well in both regimes, while $\diffmodel{1}$ smooths the spectrum in the jet regime and underestimates the kinetic energy across all wavenumbers. From Fig.\;(\ref{fig: spectra_jet_full} c, d, g, h), $\diffmodel{2}$, $\diffmodel{3}$, and $\diffmodel{4}$ accurately capture the angle-averaged spectra at all wavenumbers, including smaller scales (i.e., the tails). In the jet regime, $\diffmodel{3}$ and $\diffmodel{4}$ show slightly higher energy at the tails. $\diffmodel{1}$ is once again skewed by the data and shows steeper scaling than the target spectra; it matches the observed spectrum where available and then switches to the unconditional spectrum (stronger effect in the more complex eddy regime). For lower $Re = 10^3$ (figures not shown for brevity), the two conditional models capture the spectra better than $\diffmodel{2}$ at all wavenumbers.

\emph{Super-resolution at different down-sampling scales}: To assess performance under different super-resolution settings, we consider the harder eddy regime (here at $Re=10^3$) with down-sampling scales $\delta=2^1$ and $2^3$, in addition to $\delta=2^2$ used in the main test cases discussed above. $\diffmodel{1}$ is easy to implement and does not require any re-training since it re-uses the unconditional model with modified initial conditions that only depend on the data \eqref{eq:sdedit}. $\diffmodel{2}$ also does not require re-training for each $\delta$, but requires access to the observation operators $\obsmodel$ corresponding to each $\delta$, and their gradients \eqref{eq: DPS_mm}. $\diffmodel{3}$ and $\diffmodel{4}$ have to be re-trained from scratch for each $\delta$, limiting their flexibility, but they do not require access to the gradients of each $\obsmodel$.

\begin{figure}[h] 
    \centering   \includegraphics[width=\linewidth]{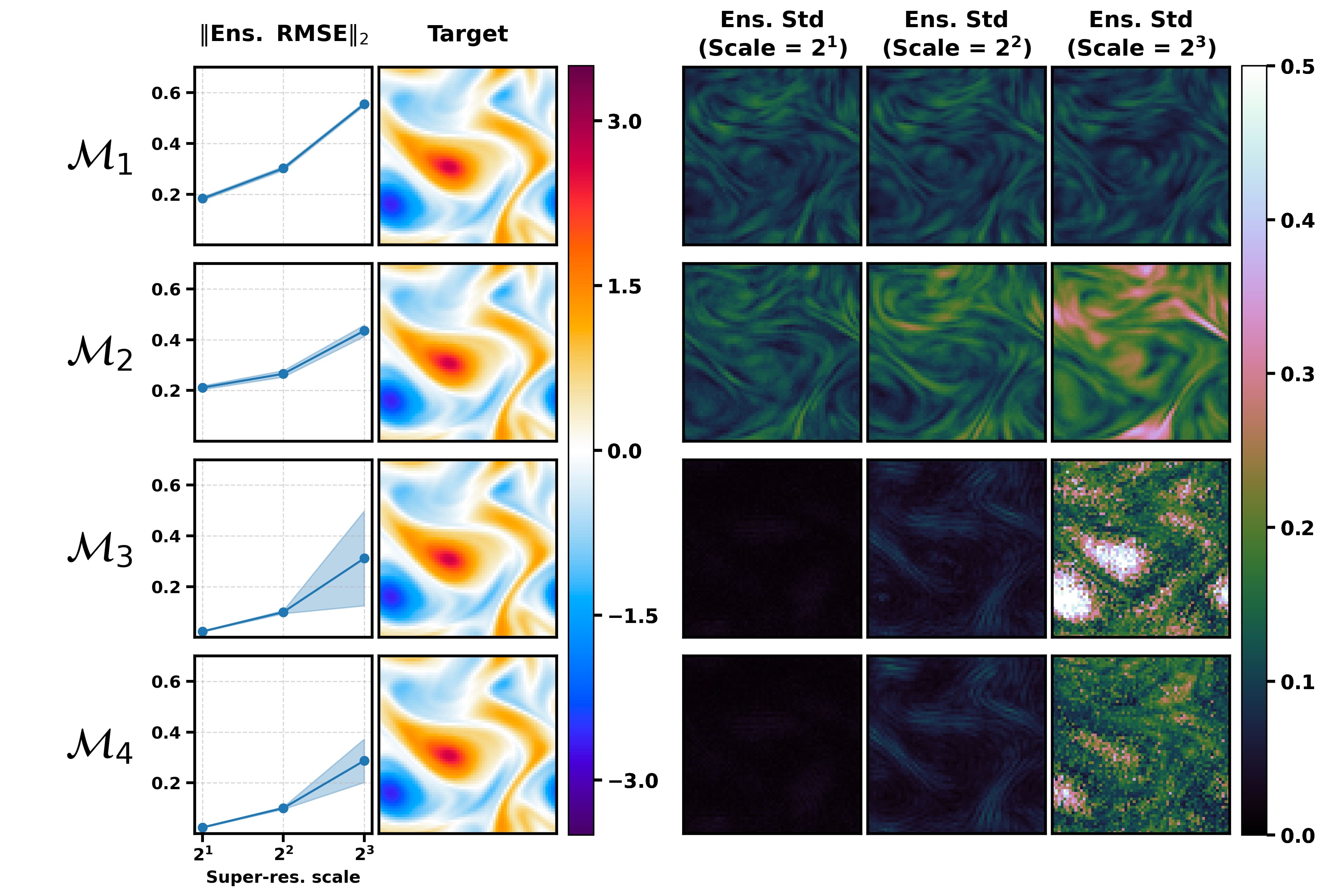}
    \caption{\small Super-resolution with the four diffusion modeling approaches at different down-sampling scales ($\delta$) in the eddy regime (Re=$10^3$) from coarse-resolution fields (Test Case 3, Table\;\ref{table: testcases}). Column 1: Norm of the relative ensemble root mean squared error (RMSE) with shaded regions showing $\pm$ one standard deviation for the reconstructed field with increasing down-sampling scales. Column 2: Target vorticity field. Columns 3-5: Ensemble standard deviations at different $\delta$. Target is identical for all models, and all $\delta$'s.
    }
    \label{fig: scales_eddy_full}
\end{figure}

Fig.\;(\ref{fig: scales_eddy_full}) shows that the ensemble RMSE norm increases with $\delta$, since coarser data contains less information for super-resolution, even for the large-scale vortices. For all methods except $\diffmodel{1}$, the standard deviation of the ensemble RMSE norm also increases with $\delta$. 
Similar trends are also observed in the corresponding ensemble standard deviation of the generated super-resolved fields. As shown for $\delta=2^2$, $\diffmodel{1}$ is inaccurate for super-resolution and generates non-physical fields, and hence its standard deviation does not vary much with $\delta$. The two conditional models $\diffmodel{3}$ and $\diffmodel{4}$ again show the smallest error norms followed by $\diffmodel{2}$, then $\diffmodel{1}$. For $\delta=2^1$, the observations contain sufficient fine-scale information for the conditional models to near-perfectly reconstruct fine-scale features (e.g., less than five percent error). For $\delta = 2^3$, $\diffmodel{3}$ and $\diffmodel{4}$ show the largest error norm standard deviations, predict the largest ensemble standard deviation fields, and their
error patterns directly relate to the target field (FF).
While expected, these error properties indicate that learning the conditional distribution between the target and the data is difficult at extremely coarse resolutions ($N_{OF} = 8 \times 8$). 

\subsection{Applications 2: Super-resolution and Inference from Coarse and Sparse, Gappy Data.} \label{sec: results-sparse}

We now discuss applications of full field super-resolution inference from coarse and sparse, gappy observations of the vorticity field, corresponding to Cases 5-8 of Table\;(\ref{table: testcases}). These synthetic data enable simple Observing System Simulation Experiments (OSSEs)  analogous to real OSSEs \cite{errico2013development} used for data assimilation of satellite-derived surface measurements such as SWOT \cite{morrow2019global} and DUACS \cite{taburet2019duacs} altimetry data. 
We avoid realistic scenarios where vorticity is inferred only from noisy, measured Sea Surface Height (SSH), and instead consider an idealized setting with direct coarse, gappy vorticity measurements. The idealized testbed enables direct model comparison without obfuscation from noise and indirect measurement, or compute-intensive training with high-resolution ocean reanalysis fields and direct satellite observations.

Table\;(\ref{table: metrics_jet_sparse}) summarizes the key quantitative skill metrics. Overall, the errors of all approaches are significantly higher (more than $2 \times$) than in the corresponding coarse-resolution cases (Table\;\ref{table: metrics_jet_full}). The eddy regime from coarse and sparse, gappy observations is the most challenging of all test cases we consider: the target fields have the most complicated distribution, while the sparse data offers the least information for super-resolution. This results in relatively lower skill metrics for all methods when compared to the jet regime cases.
We nonetheless find that trends in error metrics across models in this application remain as in Sect.\;(\ref{sec: results-full}). In the following discussion, we highlight only the key findings and differences. For some methods and test cases, we will find that the coarse, sparse data are insufficient to infer the super-resolved field. 

\begin{table}[h!]
\setlength{\tabcolsep}{3pt}
\scriptsize 
\centering
\caption{\small Comparison of the key quantitative skill metrics for super-resolution from coarse and sparse, gappy observations (Cases 5-8 of Table\;(\ref{table: testcases})).} \hspace*{-0.85in}
\begin{tabular}{l|l|cccc|cccc}
  \toprule
  \multicolumn{2}{l|}{} 
  & \multicolumn{4}{c|}{$Re=10^3$}
  & \multicolumn{4}{c}{$Re=10^4$} \\
  \midrule
  Relative Error Metric
  & Scale
  & \shortstack{$\diffmodel{1}$}
    & \shortstack{$\diffmodel{2}$}
    & \shortstack{$\diffmodel{3}$}
    & \shortstack{$\diffmodel{4}$}
    & \shortstack{$\diffmodel{1}$}
    & \shortstack{$\diffmodel{2}$}
    & \shortstack{$\diffmodel{3}$}
    & \shortstack{$\diffmodel{4}$} \\
  \toprule
 \multicolumn{2}{c}{\hspace{-1in} Jet regime} & \multicolumn{4}{c}{} & \multicolumn{4}{c}{} \\
  \toprule
  Reconstructed vorticity ($\downarrow$)  & ($\times 10^{-1}$) 
    & 6.07 $\pm$ 0.14 & 3.38 $\pm$ 0.12 & \textbf{0.45 $\pm$ 0.21} & 0.48 $\pm$ 0.13
    & 6.68 $\pm$ 0.10 & 4.35 $\pm$ 0.12 & \textbf{0.99 $\pm$ 0.04} & 1.02 $\pm$ 0.05 \\
  Ensemble Field Std. & ($\times 10^{0}$) 
    & 0.20 & 0.06 & 0.04 & 0.04
    & 0.22 & 0.10 & 0.06 & 0.06 \\
  Cycle-consistency ($\downarrow$) & ($\times 10^{-1}$) 
    & 3.58 $\pm$ 0.09 & 2.24 $\pm$ 0.07 & 0.13 $\pm$ 0.13 & \textbf{0.10 $\pm$ 0.06}
    & 3.42 $\pm$ 0.12 & 2.40 $\pm$ 0.10 & \textbf{0.10 $\pm$ 0.01} & \textbf{0.10 $\pm$ 0.01} \\
  Log-energy spectrum ($\downarrow$) & ($\times 10^{-2}$) 
    & 24.92 $\pm$ 0.30 & 12.76 $\pm$ 0.61 & \textbf{8.63 $\pm$ 2.35} & 9.56 $\pm$ 1.73
    & 18.49 $\pm$ 0.26 & 7.50 $\pm$ 0.60 & \textbf{1.64 $\pm$ 0.18} & 1.82 $\pm$ 0.30 \\
  Log-enstrophy spectrum ($\downarrow$) & ($\times 10^{-2}$) 
    & 38.41 $\pm$ 0.47 & 19.68 $\pm$ 0.95 & \textbf{13.30 $\pm$ 3.62} & 14.73 $\pm$ 2.67
    & 30.09 $\pm$ 0.43 & 12.22 $\pm$ 0.98 & \textbf{2.67 $\pm$ 0.29} & 2.95 $\pm$ 0.50 \\
  \toprule
  \multicolumn{2}{c}{\hspace{-1in} Eddy regime} & \multicolumn{4}{c}{} & \multicolumn{4}{c}{} \\
    \toprule
     Reconstructed vorticity ($\downarrow$) & ($\times 10^{-1}$) 
     & 7.75 $\pm$ 0.08 & 4.43 $\pm$ 0.53 & 2.47 $\pm$ 0.33 & \textbf{2.39 $\pm$ 0.16}
    & 6.83 $\pm$ 0.11 & 6.42 $\pm$ 0.21 & 3.13 $\pm$ 0.21 & \textbf{3.10 $\pm$ 0.18} \\
  Ensemble Field Std. & ($\times 10^{0}$)
    & 0.11 & 0.24 & 0.14 & 0.13
    & 0.16 & 0.19 & 0.17 & 0.17 \\
  Cycle-consistency ($\downarrow$) & ($\times 10^{-1}$) 
    & 3.77 $\pm$ 0.12 & 1.80 $\pm$ 0.17 & \textbf{0.16 $\pm$ 0.01} & \textbf{0.16 $\pm$ 0.01}
    & 3.08 $\pm$ 0.11 & 2.53 $\pm$ 0.14 & 0.06 $\pm$ 0.01 & \textbf{0.05 $\pm$ 0.00} \\
  Log-energy spectrum ($\downarrow$) & ($\times 10^{-2}$) 
    & 14.86 $\pm$ 1.11 & \textbf{5.82 $\pm$ 0.27} & 5.84 $\pm$ 0.43 & 6.33 $\pm$ 0.58
    & 6.12 $\pm$ 0.49 & 5.03 $\pm$ 0.56 & 3.13 $\pm$ 0.50 & \textbf{2.76 $\pm$ 0.40} \\
  Log-enstrophy spectrum ($\downarrow$) & ($\times 10^{-2}$)
    & 22.88 $\pm$ 1.72 & \textbf{8.97 $\pm$ 0.42} & 8.99 $\pm$ 0.66 & 9.75 $\pm$ 0.89
    & 9.86 $\pm$ 0.80 & 8.12 $\pm$ 0.91 & 5.06 $\pm$ 0.80 & \textbf{4.46 $\pm$ 0.65} \\
  \toprule
  \multicolumn{10}{l}{Error metrics are shown as ensemble mean $\pm$ one standard deviation; the best-performing model for each metric is highlighted in \textbf{bold}.}
\end{tabular}\label{table: metrics_jet_sparse}
\normalsize 
\end{table}

\begin{figure}[hp] \hspace*{-0.75in}
        \includegraphics[width=1.2\linewidth]{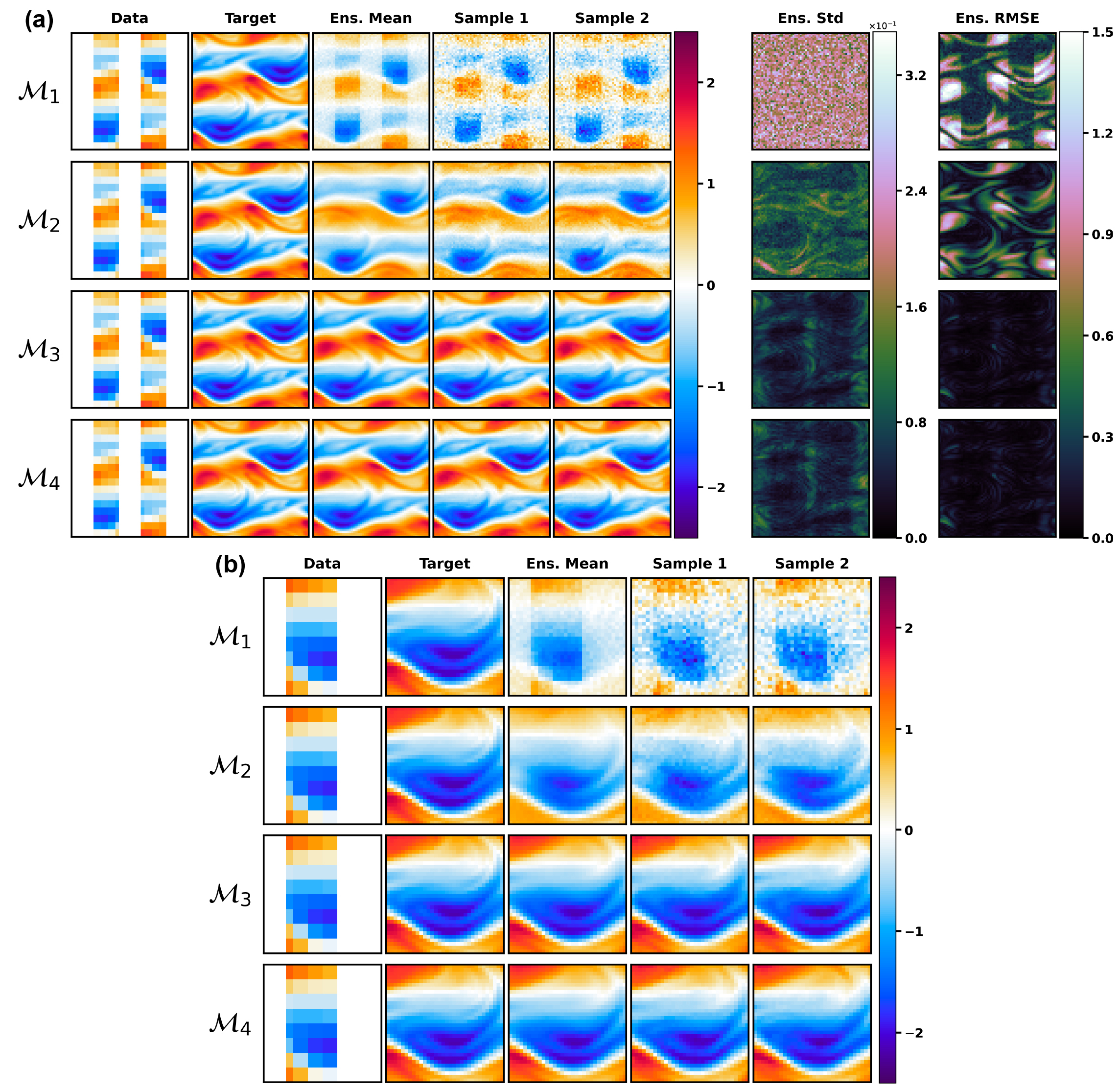}
    \caption{\small As Fig.\;(\ref{fig: fields_jet_full}), but for super-resolution and inference in the jet regime (Re=$10^4$) from coarse and sparse, gappy observations (Test Case 6, Table\;\ref{table: testcases}).}
    \label{fig: fields_jet_sparse}
\end{figure}

\begin{figure}[hp] \hspace*{-0.75in}
        \includegraphics[width=1.2\linewidth]{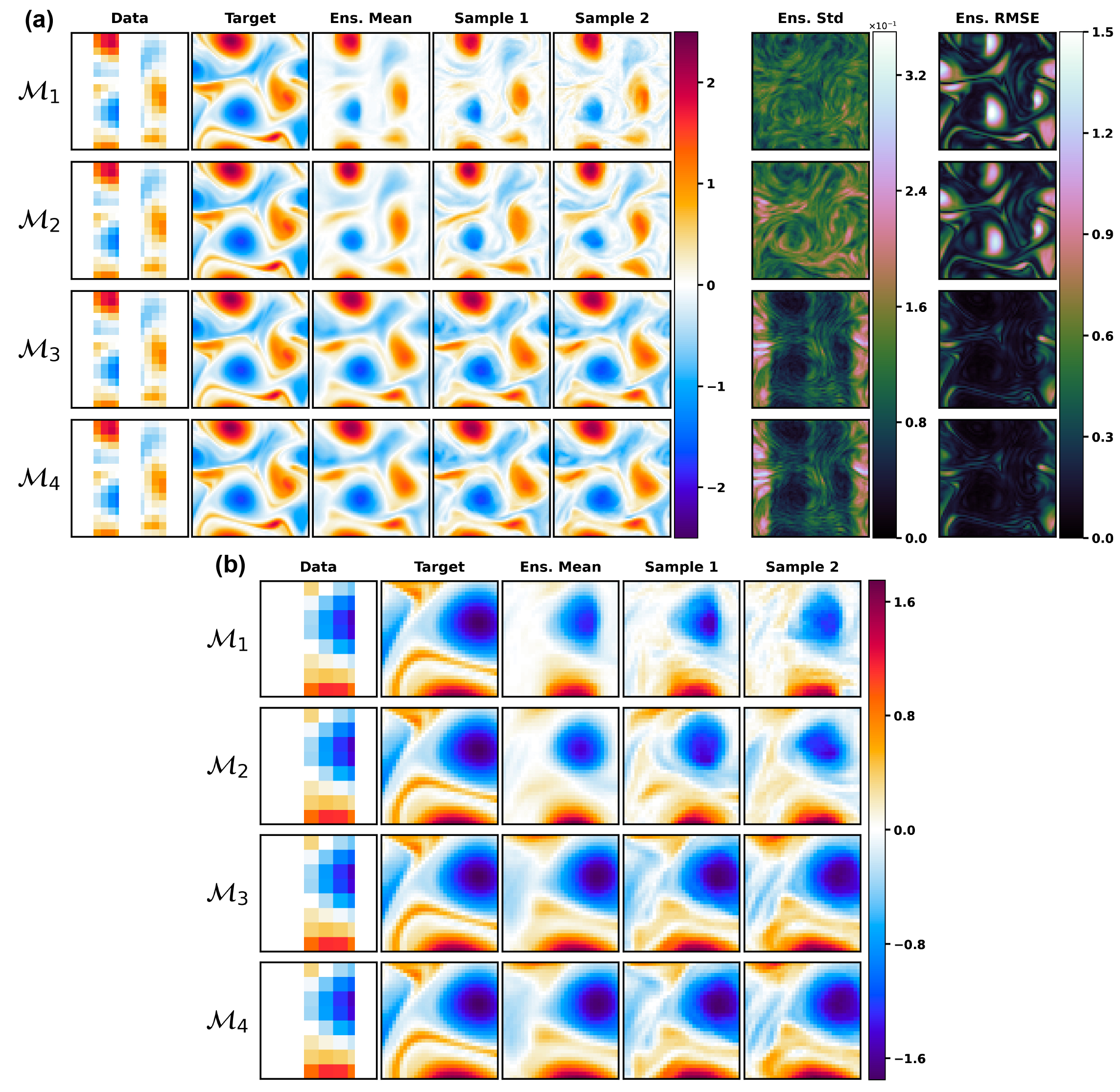}
    \caption{\small As Fig.\;(\ref{fig: fields_eddy_full}), but for super-resolution and inference in the eddy regime (Re=$10^4$) from coarse and sparse, gappy observations (Test Case 8, Table\;\ref{table: testcases}).}
    \label{fig: fields_eddy_sparse}
\end{figure}

\emph{Super-resolved vorticity fields}:  
From Figs.\;(\ref{fig: fields_jet_sparse}) and (\ref{fig: fields_eddy_sparse}), $\diffmodel{1}$ fails to generate physically valid fields, struggling to capture even the large-scale zonal jets and vortex cores. While SDEdit has shown promise in stroke-based editing and synthesis in computer vision, it performs poorly when initialized with gappy data with low-to-moderate noise levels. In fact, $\diffmodel{1}$ does not provide any theoretical guarantees of convergence to the posterior. Though initializing with higher levels of noise could lead to more realistic fields, this approach would be akin to unconditional generation \cite{meng2021sdedit}, which results in low cycle consistency and larger relative errors due to phase incoherence (Sect.\;\ref{sec: results-sensitivity}). $\diffmodel{2}$ also does not infer vortex core boundaries well and generates filaments with unphysical noise. While $\diffmodel{2}$ can estimate the state given both full-coarse fields and sparse, gappy, high-resolution observations \cite{martin2025generative}, here it underperforms with the limited observations.
We also note that in the eddy regime at $Re=10^4$, the sparse, gappy observations are not sufficiently informative to infer all finer-scale structures in the data gaps, leading to greater ensemble variability of the reconstructed filaments.


\emph{Uncertainty quantification}:
For all models, in both regimes, the ensemble RMSE is highest in the unobserved regions, followed by 
observed boundary regions. We analyze the RMSE against the variability of the target in Sect.\;(\ref{sec:results-sparse-analysis}).
For $\diffmodel{3}$ and $\diffmodel{4}$, once again, the ensemble standard deviation 
spatially follows the ensemble RMSE;
these models can identify low-accuracy regions even in this challenging case
\cite{souza2025surface}.

\begin{figure}[]\hspace*{-1in}
        \includegraphics[width=1.4\linewidth]{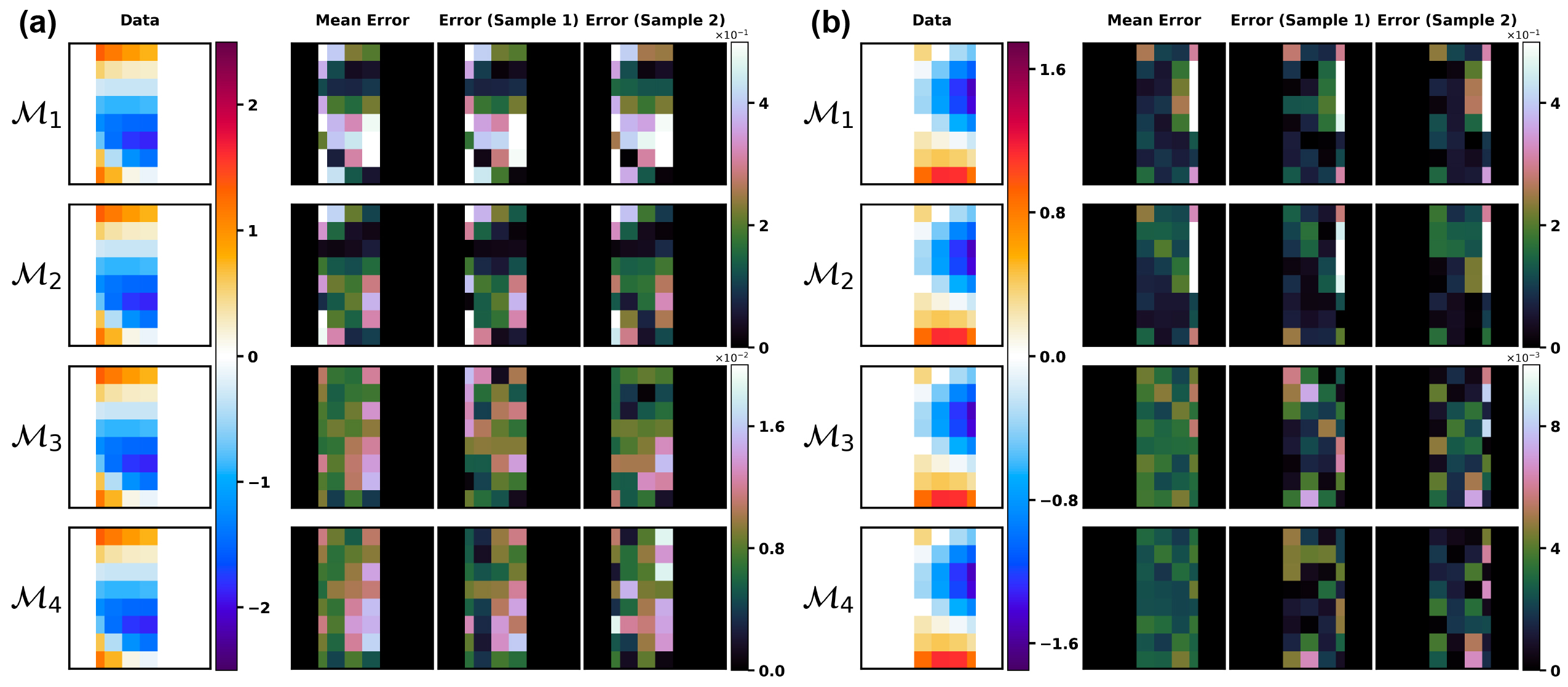} 
    \caption{\small As Fig.\;(\ref{fig: cons_jet_full}), but with zoomed-in views of the (a) Upper-right quadrant for super-resolution in the jet regime (Re=$10^4$) from coarse and sparse, gappy observations (Test Case 6, Table\;\ref{table: testcases}) (b) Lower-left quadrant for super-resolution in the eddy regime (Re=$10^4$) from coarse and sparse, gappy observations (Test Case 8, Table\;\ref{table: testcases}).
    }\label{fig: cons_jet_sparse}
\end{figure}

\emph{Cycle-consistency}: 
Remarkably, $\diffmodel{3}$ and $\diffmodel{4}$ are cycle-consistent even in this sparse observation application (Fig.\;\ref{fig: cons_jet_sparse}). So direct conditional information can enforce even out-of-distribution cycle-consistency through training. We investigate conditioning further in Sect.\;(\ref{sec:results-sparse-analysis})

\begin{figure}[]\hspace*{-1in}
    \includegraphics[width=1.4\linewidth]{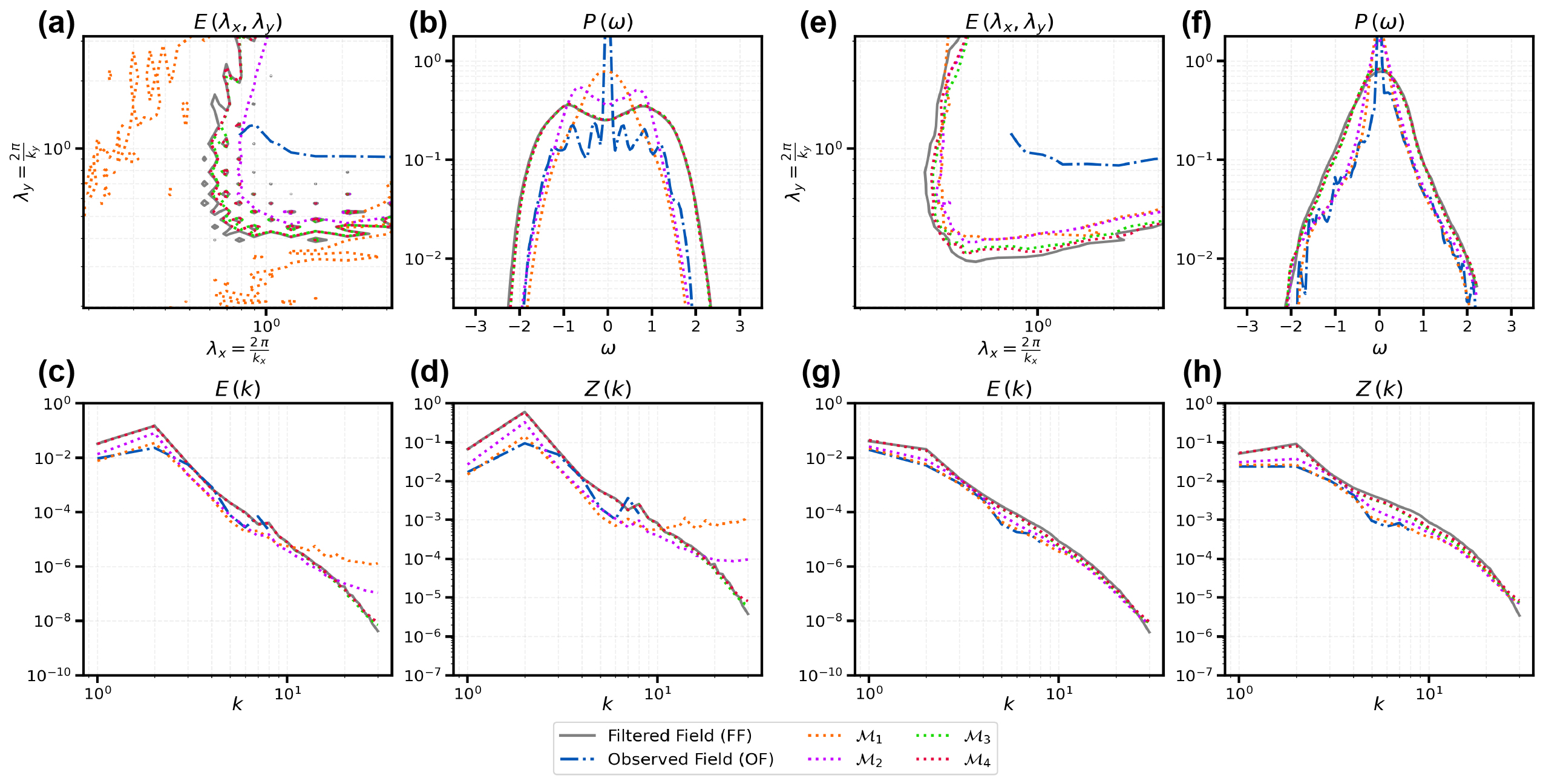}
     \caption{\small As Fig.\;(\ref{fig: spectra_jet_full}), but for super-resolution and inference from coarse and sparse, gappy observations in the (a-d) jet regime (Re=$10^4$) (Test Case 6, Table\;\ref{table: testcases}) and (e-h) eddy regime (Re=$10^4$) (Test Case 8, Table\;\ref{table: testcases}).}
    \label{fig: spectra_jet_sparse}
\end{figure}

\emph{Time-averaged 2D turbulence statistics}: Unlike the super-resolution applications, $\diffmodel{1}$ completely misses the bimodal distribution in the jet regime (Fig.\;\ref{fig: spectra_jet_sparse} b).
$\diffmodel{2}$ captures the bimodality but significantly underestimates large values of the vorticity (Figs.\;\ref{fig: spectra_jet_sparse} b, f). Similarly, $\diffmodel{1}$ completely mispredicts the shape of the spectrum contours in the jet regime (Fig.\;\ref{fig: spectra_jet_sparse} a) while $\diffmodel{2}$ underestimates the energy at all wavenumbers in the eddy regime (Fig.\;\ref{fig: spectra_jet_sparse} e). For the kinetic energy and enstrophy spectra (Figs.\;\ref{fig: spectra_jet_sparse} c, d, g, h), $\diffmodel{1}$ and $\diffmodel{2}$ shift the spectra significantly to smaller scales since they generate fields with significant fine-scale noisy artifacts.

\begin{figure}[]\hspace*{-0.5in}
    \includegraphics[width=1.2\linewidth]{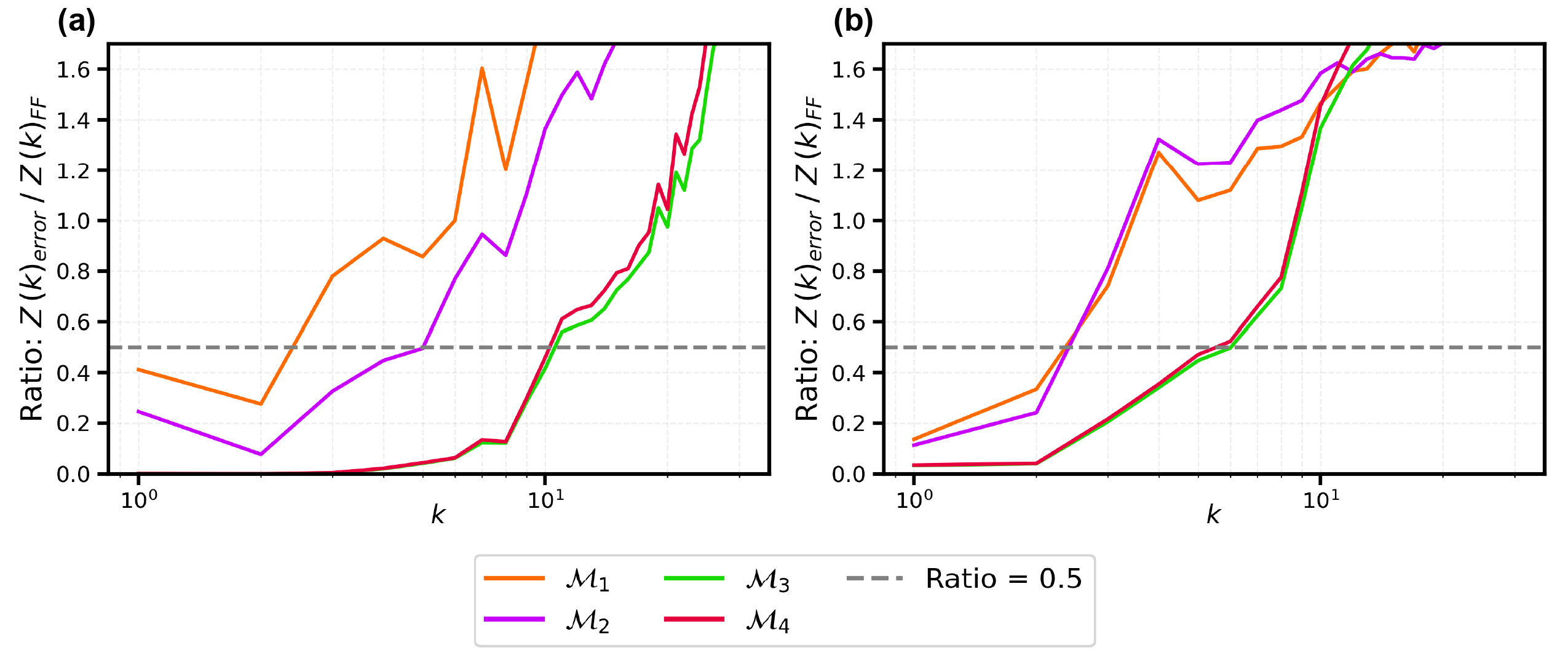} 
     \caption{\small Time-averaged noise-to-signal ratios of the spectrum for super-resolution and inference from coarse and sparse, gappy observations in the (a) jet regime (Re=$10^4$) (Test Case 6, Table\;\ref{table: testcases}) and (b) eddy regime (Re=$10^4$) (Test Case 8, Table\;\ref{table: testcases}).}
    \label{fig: effective_resolution_sparse}
\end{figure}

Finally, we compute the time-averaged noise-to-signal ratio of the spectrum \eqref{eq: NSR_ratio}. A threshold ratio of 0.5 is used to estimate the mean effective resolution \cite{ballarotta2019resolutions}. Compared to the
prediction-to-signal spectral magnitude ratio
\cite{chelton2011global}, the effective resolution is a stricter measure since it takes into account both the amplitude
and phase coherence of resolved features. 
Fig.\;(\ref{fig: effective_resolution_sparse}) shows that
the unconditional models' effective resolution is significantly degraded compared to that of the conditional models. The conditional effective resolution seems to match the statistical limit where ensemble variability (Figs. \ref{fig: fields_jet_sparse} and \ref{fig: fields_eddy_sparse}) drives error, across both the super-resolution and inference applications. 
The conditional spectra show appropriate scaling, but with large noise-to-signal ratios beyond the effective resolution, suggesting that these models may overestimate eddy and filament magnitudes.
The loss of phase coherence could be due to the probabilistic nature of diffusion models, since each ensemble member represents a plausible vorticity field.

\subsubsection{Propagation of Observational Information} \label{sec:results-sparse-analysis}

We now analyze how information from the coarse and sparse, gappy observations is propagated to the unobserved regions during the reverse stochastic processes \eqref{eq:ito_reverse} and \eqref{eq:ito_reverse_conditional}. For linear inverse problems, where $\obsmodel = H \in \mathbb{R}^{\hrdim} \times \mathbb{R}^{\lrdim}$, \citeA{dou2024diffusion} showed an equivalence between posterior sampling from the conditional distribution and Bayesian filtering. Under this \add{Gaussian} approximation, the Kalman filter \cite{kalman1960new} provides an optimal solution to the filtering problem, where observational information is propagated through the covariance terms appearing in the Kalman gain, namely the covariance of the forecast error and covariance of the observational error. In the ensemble setting, the forecast error covariance is estimated directly from the ensemble members \cite{lermusiaux_robinson_MWR1999, evensen2003ensemble}. Similarly, a term analogous to the \add{posterior} forecast error covariance, \add{$\text{Cov}[x_0|x_t,y]$ (\ref{eq: cov_diffusion})}, can be derived analytically for score-based diffusion models \cite{boys2023tweedie} \add{using the conditional reverse process score} \eqref{eq: Tweedie conditional}\change{ as}{.}
\begin{equation}\label{eq: cov_diffusion}
    \text{Cov}[x_0|x_t,y] = \frac{\sigma^2_t}{\mu_t^2}\bigg[\mathcal{I}_{\hrdim} + \sigma_t^2 \nabla^2_{x_t}\bigg(\log p_t(x_t|y)\bigg)\bigg]
\end{equation}
\change{although}{Generally,} it is computationally intractable to compute the Hessian of the conditional score\add{, $\nabla^2_{x_t}\bigg(\log p_t(x_t|y)\bigg)$} \cite{rissanen2024free}.

\begin{figure}[h!] \hspace*{-0.35in}
    \includegraphics[width=1\linewidth]{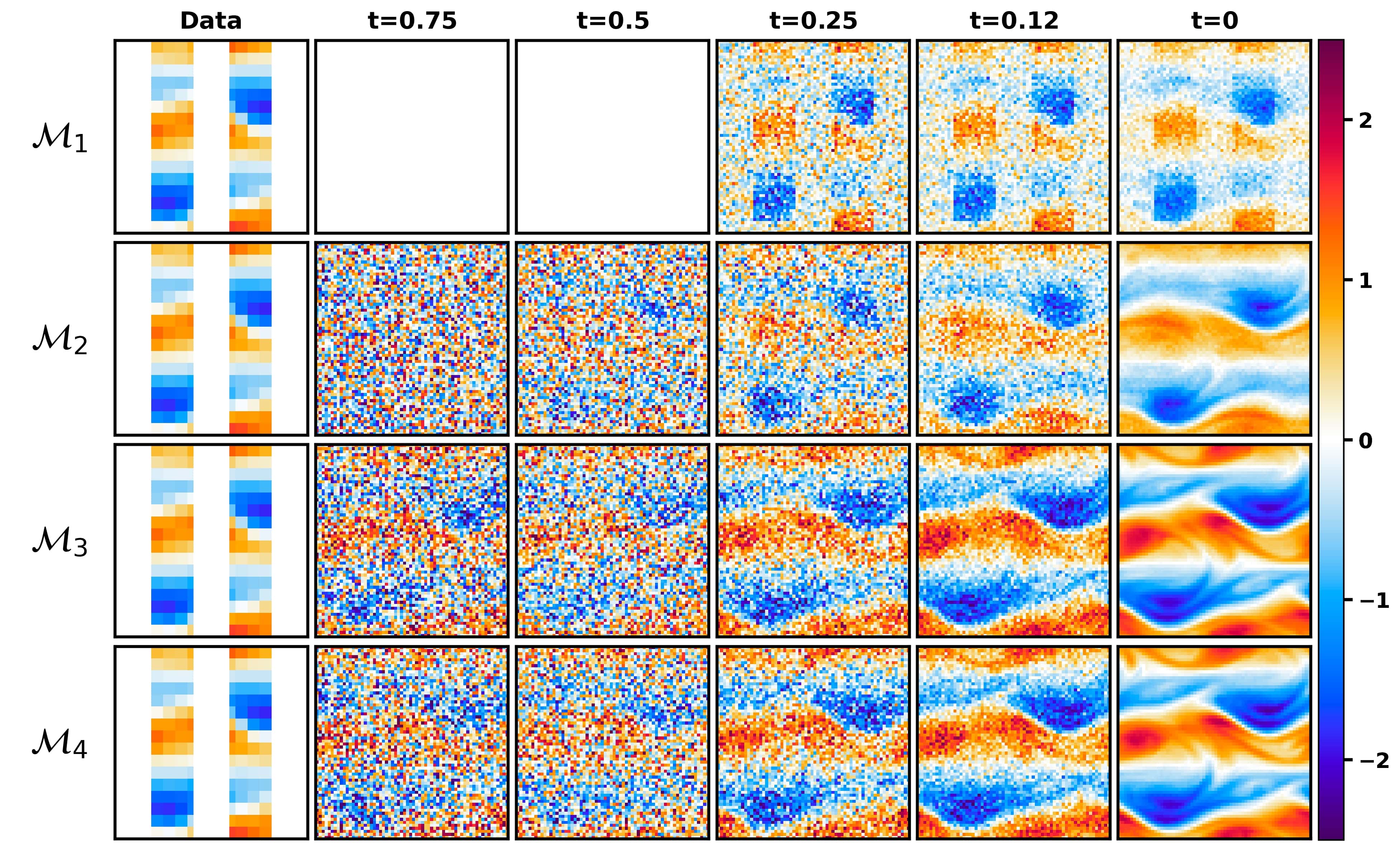}
    \caption{\small Evolution of vorticity fields during the reverse diffusion process from coarse and sparse, gappy observations in the jet regime (Re=$10^4$, Test Case 6, Table\;\ref{table: testcases}).}
    \label{fig: obs_rev_sde}
\end{figure}

Fig.\;(\ref{fig: obs_rev_sde}) shows the evolution of the reverse diffusion process for Test Case 6 (jet regime). For $\diffmodel{1}$, we find that its initialization at $t_i=0.25$ preserves the large-scale features of the data. However, the unconditional reverse process \eqref{eq:ito_reverse} does not propagate this information to the unobserved regions, since it only utilizes the unconditional score, which is independent of observations, leading to poor reconstruction of large and fine-scale features. For $\diffmodel{2}$, we find some noisy but recognizable zonal jets at $t=0.5$, which indicates the propagation of observational information to unobserved regions, albeit gradually and more diffuse compared to the conditional models. This could be because $\diffmodel{2}$ approximates the true covariance \eqref{eq: cov_diffusion} with just $\frac{\sigma_t^2}{\mu_t^2}\mathcal{I}_{\hrdim}$ in unobserved regions \cite{rozet2023score}. For the two conditional models, we observe large, noisy zonal jets at much earlier diffusion pseudo-times ($t=0.75$), with sharper jets visible at $t=0.5$. This effect is also observed in the spectra at intermediate diffusion pseudo-times (figures not shown), where the conditional models show higher energies at larger scales much earlier compared to the unconditional models. This strong conditioning could explain their cycle-consistency (Fig.\;\ref{fig: cons_jet_sparse}).

\begin{figure}[]\hspace*{-0.35in}
    \includegraphics[width=1\linewidth]{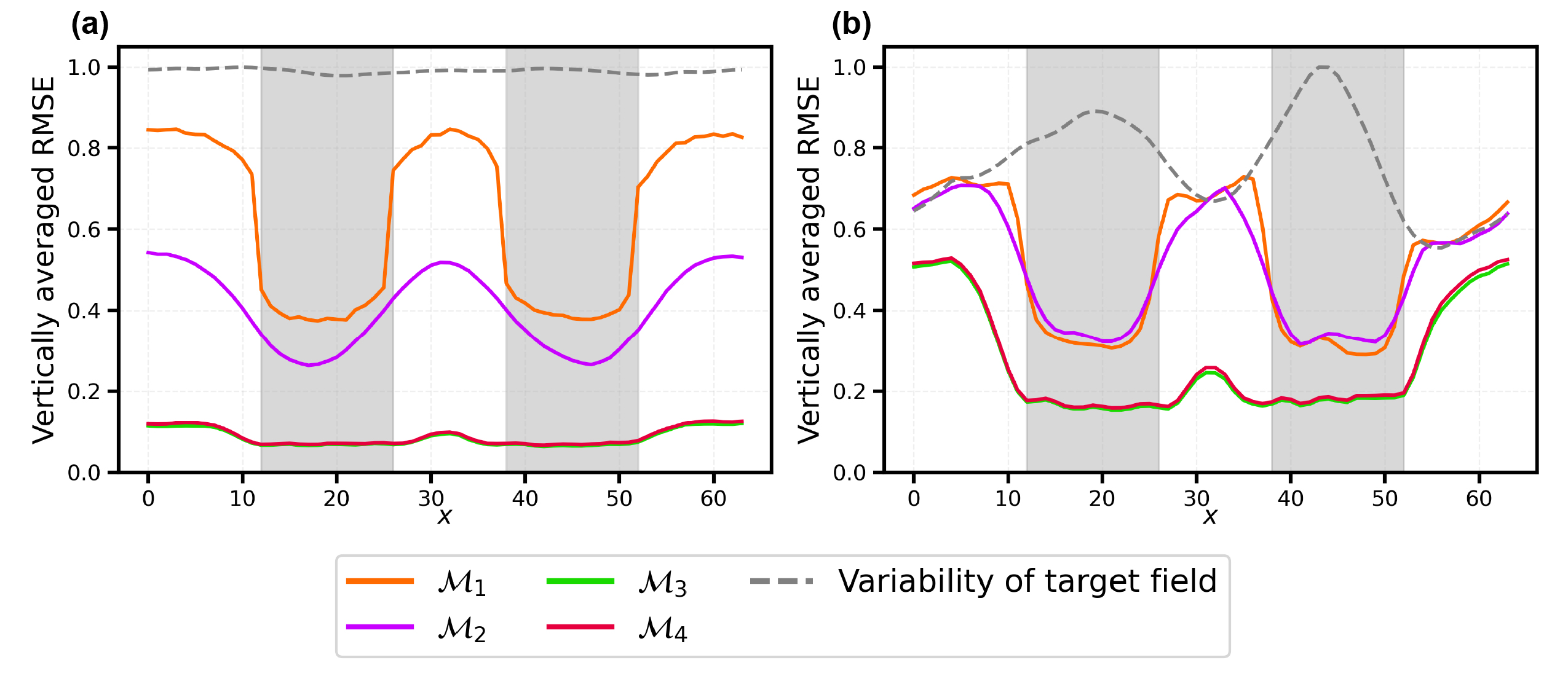}
     \caption{\small Time and vertically averaged RMSE compared to the variability of the target for super-resolution and inference from coarse and sparse, gappy observations 
     (data regions shaded in gray) in the (a) jet regime (Re=$10^4$) (Test Case 6, Table\;\ref{table: testcases}) and (b) eddy regime (Re=$10^4$) (Test Case 8, Table\;\ref{table: testcases}).}
    \label{fig: vertical_error_sparse}
\end{figure}

Fig.\;(\ref{fig: vertical_error_sparse}) shows the variation of the vertically averaged RMSE of the four diffusion models with distance from observed regions. We observe similar spatial trends in RMSE as in Figs.\;(\ref{fig: fields_jet_sparse}) and  (\ref{fig: fields_eddy_sparse}). However, in the harder eddy regime, the RMSE of the two unconditional methods approaches the variability of the target field in the middle of the unobserved regions, indicating that these models do not propagate information there effectively. For $\diffmodel{2}$, this can be explained by analyzing \eqref{eq: DPS_final} in unobserved regions where the measurement matching term \add{$\nabla_{x_t}$}($\norm{y - \obsmodel(\mathbb{E}[X_0|X_t=x_t])})_\text{unobserved}\to 0$
\add{(In this case, the innovation vector $y - \obsmodel(\mathbb{E}[X_0|X_t=x_t])_\text{unobserved}$ also tends to 0 everywhere away from data).}
Hence, we obtain \eqref{eq: DPS_unobserved}.
\begin{equation} \label{eq: DPS_unobserved}
    \nabla_{x_t}\bigg(\log p_t(x_t|y)\bigg)_{\diffmodel{2}, \text{unobserved}} \simeq \nabla_{x_t}\bigg(\log p_t(x_t)\bigg)
\end{equation}
%
Therefore, unobserved regions can lose guidance, leading to unconditional vorticity field generation. As in traditional splitting schemes, the reverse process can only propagate information through the interaction between the unconditional score and the measurement matching term, with the observation operator's physical bandwidth (e.g., width of tracks) determining the propagation rate.
Only in the continuous-time limit of infinite diffusion steps can $\diffmodel{2}$ theoretically approximate a conditional model.
In practice, the reverse diffusion process is run with Langevin Monte Carlo corrections (\ref{sec: training}), which allow global interaction, improving information propagation (as seen in the jet regime in Fig.\;(\ref{fig: fields_jet_sparse})).
We also studied the correlation between observations and unobserved regions (figure not shown), with the conditional models exhibiting stronger statistical convergence than the unconditional ones.

\subsection{Sensitivity Studies: Effect of Tuning Parameters}\label{sec: results-sensitivity}

Finally, we study the sensitivity of the four models to their tuning parameters and provide their optimal values. All above results used optimally tuned intermediate times (for $\diffmodel{1}$) and guidance strengths (for $\diffmodel{2}, \diffmodel{4}$). $\diffmodel{3}$ does not require any tuning; it only relies on training. The choice of intermediate time and guidance strength greatly controls the quality of the generated super-resolved samples, and trades off fidelity (i.e., physical plausibility and sharpness of the generated fields) for consistency (i.e., cycle-consistency with the observations or data). Since the denoiser neural networks, \eqref{eq: training_obj_uncond} and \eqref{eq: training_obj_cond}, are trained independently of these parameters, their weights are unaffected by this fine-tuning. This allows for great flexibility, as the intermediate time and guidance strengths can be optimized and re-tuned post-training for each task (application), or even for each snapshot within an application \cite{chung2023diffusion}. 

An alternate approach to improve the fidelity of the generated samples is to optimize the initial Gaussian noise used for the reverse process \cite{qi2024not}. However, this is computationally expensive, and hence, we use a fixed seed to generate Gaussian initial conditions and focus on fine-tuning only the intermediate time and guidance strengths. We optimize over all snapshots of super-resolution by sweeping over a range of values and choosing the parameter value that minimizes the $L_2$ norm of the error of the super-resolved fields. For brevity, we discuss results below for super-resolution in the jet regime (Re=$10^4$) from coarse-resolution field observations. Similar trends hold for all other applications and test cases.

\subsubsection{$\diffmodel{1}$:  Modifying the Initial Condition (SDEdit)}
\begin{figure}[] \hspace*{-0.35in}
    \centering
    \includegraphics[width=1.2\linewidth]{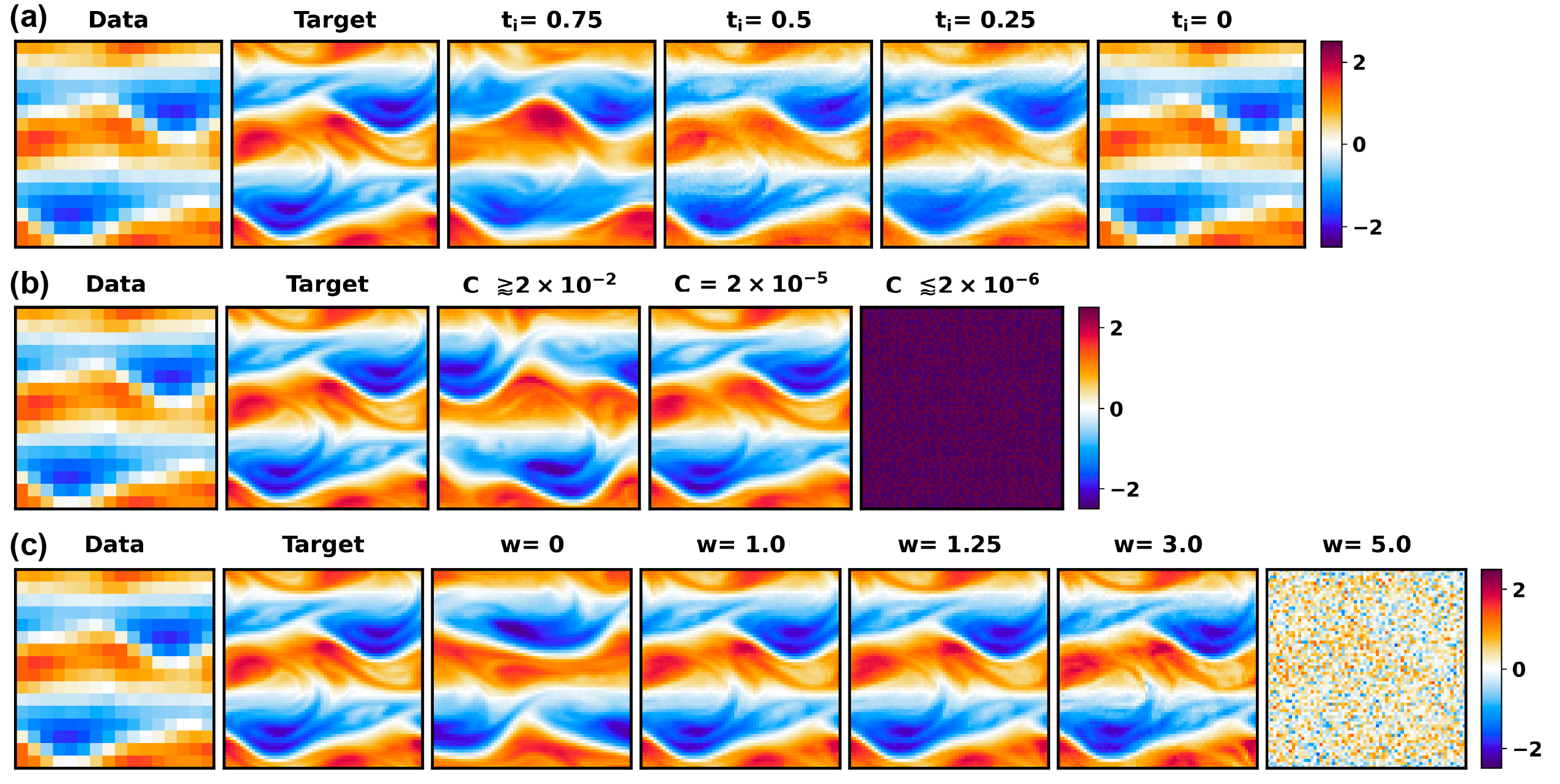}
    \caption{\small Sensitivity of (a) $\diffmodel{1}$ (guidance by modifying the initial condition) to the intermediate pseudo-time $\sdedittime$, (b) $\diffmodel{2}$ (guidance by modifying the score) to the guidance strength $C$, and (c) $\diffmodel{4}$ (classifier-free guidance) to the guidance strength $w$.}
    \label{fig: sens_sde}
\end{figure}

The main parameter to be tuned is the intermediate time $\sdedittime$. Fig.\;(\ref{fig: sens_sde} a) shows that choices of $\sdedittime$ close to $0$ lead to unphysical super-resolved fields. Since $\sdedittime = 0$ corresponds to the final diffusion pseudo-timestep, the model does not get to denoise the input field (data). On the other hand, choices of $\sdedittime$ close to $T=1$ ($\sdedittime\gtrapprox 0.75$) lead to fields that lose track of large-scale features of the data, and hence generate non-cycle-consistent super-resolved fields \cite{meng2021sdedit}. Setting $0.1 \lessapprox \sdedittime \lessapprox 0.6$ generates relatively cycle-consistent super-resolved fields, and we obtain the smallest relative error at $\sdedittime=0.25$.

\subsubsection{$\diffmodel{2}$: Modifying the Score (Diffusion Posterior Sampling)}
The main parameter to be tuned here is the guidance strength, $C$. From \eqref{eq: DPS_scaling} and \eqref{eq: DPS_final}, using a value of $C \to \infty$ is equivalent to simply utilizing an unconditional model with no guidance. Fig.\;(\ref{fig: sens_sde} b) shows that setting $C\gtrapprox2\times 10^{-2}$ generates non-cycle-consistent fields. Using $ C\to 0$ corresponds to extremely strong guidance, which leads to over-saturation and unphysical noisy artifacts \cite{chung2023diffusion}, as seen in the figure for $C \lessapprox 2 \times 10^{-6}$. Here, the optimal value of $C$ was found to be $2\times 10^{-5}$, which sufficiently trades off fidelity for consistency. 

\subsubsection{$\diffmodel{3}$: Vanilla Conditional Model}
For this approach, the model directly learns the conditional score through training, and hence, there is no additional parameter to fine-tune.

\subsubsection{$\diffmodel{4}$: Classifier-Free Guidance}
The main parameter to be tuned is the guidance strength, $w$. Setting $w=0$ corresponds to sampling only from the unconditional model, while $w=1$ corresponds to sampling only from the conditional model \cite{ho2021classifier}, as seen in Fig.\;(\ref{fig: sens_sde} c). Typically, the value of $w$ is set to be greater than $1$. We obtain the smallest relative error at $w = 1.25$. For our applications, a value of $w \gtrapprox 2$ leads to the generation of spurious high-frequency features. This can be explained by the fact that classifier-free guidance performs a linear combination of the conditional and unconditional scores, and therefore, fields are generated from an approximated distribution that may not be physically meaningful \cite{karras2024guiding}. Using a very large value of $w$ leads to unphysical noisy artifacts as shown in the last column.

\section{Conclusions and Discussion}\label{sec: conclusions}

We investigated super-resolution and inference of quasi-geostrophic turbulence under the $\beta$-plane approximation with four generative diffusion modeling approaches -- two guided unconditional models and two conditional models. 
The conditional models, vanilla
($\diffmodel{3}$) and classifier-free guidance ($\diffmodel{4}$), are stable and generate accurate ensembles of super-resolved vorticity fields across all dynamical regimes (eddy and jet) and Reynolds numbers ($Re = 10^3, 10^4$). Moreover, they predict correct turbulence statistical quantities such as the 1D probability distribution function and kinetic energy spectra and enstrophy spectra (including the tails), and generalize to unseen sparse and gappy coarse observations. 
These generated ensembles are remarkably cycle-consistent with the coarse observations, despite not being explicitly optimized for during training. These two approaches show promise for deployment in geophysical inference applications where capturing accurate fine-scale features is critical, especially when the observation operator is nonlinear or indirect. Though these advantages come at the cost of application-specific re-training, they do not require knowledge of the observation operators for training or generation. On the other hand, if computational resources do not allow for re-training, guidance by modifying the score ($\diffmodel{2}$) trades off fidelity (sharpness) for cycle-consistency and generates smoothed fields in applications with coarse observations but requires access to gradients of the observation operator at each diffusion pseudo-timestep. However, $\diffmodel{2}$ does not propagate observational information effectively to unobserved regions, especially when the observation operator $\obsmodel$ is sparse and not global. Although easily implemented and computationally cheap, guidance by modifying initial conditions ($\diffmodel{1}$) can not produce correct statistics, and 
fails in challenging applications with sparse and gappy data. 

Computationally, our diffusion models are fast super-resolution and inference schemes ($\diffmodel{1}$ takes $\sim 0.1\,s$, $\diffmodel{2}$ $\sim  2\,s$, $\diffmodel{3}$ $\sim  0.3\,s$, and $\diffmodel{4}$ $\sim  0.7\,s$ on an NVIDIA A6000 GPU, computed by averaging 30 independent evaluations). $\diffmodel{2}$ takes significantly longer than the others as it requires the computation of additional gradients to evaluate the measurement matching term. A major advantage of our super-resolution framework is that it does not require auto-differentiability or access to numerical solvers, and only needs either paired high- and coarse-resolution fields or 
access to the observation operator.

Considering the future, our results with sparse and gappy observation systems show that, for conditional diffusion models, the ensemble standard deviations are good estimators of the actual errors of predictions.
This capability could be leveraged for \add{data assimilation and} adaptive sampling, i.e, identify the most optimal types and locations of observations for inference \cite{lermusiaux_PhysD2007,lermusiaux_et_al_TheSea2017}. 
Another opportunity is the super-resolution of real dynamics such as ring formation, eddy separation, and submesoscale interactions for the Gulf Stream system \cite{chassignet2008gulf,mensa2013seasonality,gula2019submesoscale}, Loop Current system 
\cite{oey2005loop,bracco2019mesoscale,nickerson2022evolution}, or other highly-dynamic mesoscale systems.
Related possibilities include the diagnosis of submesoscales \cite{zhang2024submesoscale,archer2025wide} and the inference of subsurface fields \cite{lermusiaux_et_al_Oceanog2011,klemas2014subsurface,moore_et_al_FMS2019} using sparse, gappy, coarse or indirect observations. 
However, several major limitations must be overcome for all these real applications to materialize. First, training requires access to high-resolution numerical simulations and reanalyses 
\cite{moore_et_al_FMS2019,storto2019ocean} and access to multivariate or multi-modal observations \cite{lermusiaux_et_al_oceans2002,qu2024deep}.
Second, our diffusion architecture requires extensions such as multiscale patch-wise modeling \cite{brenowitz2025climate} or modeling in a latent space \cite{rombach2022high} to handle multivariate dynamics and larger ocean and atmosphere regions. Third, novel robust evaluation metrics \cite{ballarotta2019resolutions} are likely necessary because the magnitudes and phases of the generated fields, the quantities derived from them, and the multivariate relations among them may differ from those of real fields. Ultimately, high-resolution real-world observational data and dedicated observing campaigns would be useful to evaluate and compare the performance of the different generative models, even if the training data were themselves provided by high-resolution simulations.
Extensions to spatio-temporal super-resolution and inference can be investigated through the training of guided video diffusion \cite{ho2022video} and hybrid-coupling with other stochastic numerical models \cite{lu_lermusiaux_PhysD2021} or neural surrogates \cite{rajagopal_et_al_Oceans2023}. 
Our generative diffusion is also promising for stochastic data-driven closure modeling with uncertainty estimates, e.g., for stochastic subgrid forcing \cite{lermusiaux_JCP2006,
zanna2017scale, perezhogin2023generative}.


\acknowledgments
We thank C.\;Daskalakis and G.\;Daras for insightful discussions on diffusion modeling. We thank the members of our ML-SCOPE MURI team and of our MSEAS group, including V.A.\;Rodriguez, F.M.\;Benfenati, and A.K.\;Saravanakumar, for discussions on quasi-geostrophic turbulence, statistics, and figures. We also thank the editor Dr.\ Stephen Griffies, associate editor Dr.\ Ian Grooms, and two anonymous reviewers for their constructive feedback to improve this manuscript. We are grateful to the Office of Naval Research for partial support under grant N00014-20-1-2023 (MURI ML-SCOPE) and N00014-19-1-2693 (IN-BDA) to the Massachusetts Institute of Technology. ANSB was partially supported by an MIT Mechanical Engineering MathWorks Fellowship.

\section*{Open Research}
Codes for quasi-geostrophic simulations are available via the MIT license (\url{https://github.com/ananthu545/qg-2d}) and archived at \url{https://doi.org/10.5281/zenodo.17282193} \cite{suresh_babu_et_al_2025c}. Data from the simulations are available at Zenodo via \url{https://doi.org/10.5281/zenodo.15742145} under a Creative Commons Attribution 4.0 International license \cite{suresh_babu_et_al_2025a}.
Codes for the diffusion model training and generation, and trained weights are preserved at \url{https://doi.org/10.5281/zenodo.15750243}, available via the MIT license and developed openly at \url{https://github.com/ananthu545/quasi-geostrophic-beta-plane-super-resolution} \cite{suresh_babu_et_al_2025b}.


\appendix

\section{Diffusion Model Architecture and Training}\label{sec: training}

To train the unconditional diffusion model, we utilize a U-Net architecture \cite{ronneberger2015u}.  We use channels of size $[32, 64, 128, 256]$. Following \citeA{nichol2021improved, rozet2023score}, we use the following cosine noise schedule, 
\begin{eqnarray}
    \omega = \arccos{\sqrt{10^{-3}}}\\
    \mu(t) = cos^2(\omega t) \\ 
    \sigma(t) = \sqrt{1 - \mu(t)^2} 
\end{eqnarray}
For training and sampling, we utilize $N_{diff} =64$ uniform timesteps with diffusion time $0<t<1$ with $N_{corr}=2$ Langevin Monte Carlo correction steps per diffusion pseudo-time-step, with correction size of $0.3$. Hence, for sampling, the unconditional diffusion model requires $N_{diff} \times (1+N_{corr}) = 64 \times 3 = 192$ forward passes of the trained neural network. All the vorticity fields were normalized to have zero mean and unit variance for training. The unconditional models were trained separately for each combination of regime and Reynolds number, but were shared across the two types of observations (Table \ref{table: testcases}). Transfer learning between regimes and Reynolds numbers and its mechanism are areas of active research in subgrid modeling \cite{ross2023benchmarking, subel2023explaining}, but are not explored in this work.

To train the conditional model, we follow the same procedure as for the unconditional model, but incorporate conditioning by concatenating the interpolated low-resolution field to the noisy field at each diffusion time-step following \citeA{saharia2022image}. 
We reiterate that conditional models are application-specific and were trained separately for each main test case (Table \ref{table: testcases}) and down-sampling scale ($\delta$). 
We utilize the AdamW optimizer \cite{loshchilov2017decoupled} with a cosine learning rate scheduler \cite{loshchilov2016sgdr} with learning rate = $0.0002$. We use a batch size of 64 and train for 250 epochs. We use NVIDIA A6000 GPUs for training, with each epoch taking $\sim 15$ minutes of training on a single GPU. We experimented by adding attention to the bottleneck layer of the U-Net and providing the diffusion pseudo-time $t$ as an input to the network, but did not observe noticeable performance improvement.

\bibliography{introduction,methods,results,refs_misc,mseas}

\end{document}